\documentclass[preprint]{elsarticle}
\usepackage[margin=1.3in]{geometry}
\usepackage{enumitem} 
\usepackage{graphicx}
\usepackage{subcaption}  
\usepackage{url}
\usepackage{amssymb} 
\usepackage{float}

\begin{document}

\begin{frontmatter}

\title{Leveraging inter-firm influence in the diffusion of energy efficiency technologies: An agent-based model}

\author[rvt,cot,focal]{Yingying Shi}
\author[rvt,focal]{Yongchao Zeng\corref{cor1}}
\ead{zengyc@bit.edu.cn}
\author[rvt,rvt1]{Jean Engo}
\author[rvt,cot]{Botang Han}
\author[els,focal]{Yang Li}
\author[focal]{Ralph T. Muehleisen}

\cortext[cor1]{Corresponding author at: School of Management and Economics, Beijing Institute of Technology, 5 South Zhongguancun Street, Haidian District, Beijing 100081, China}

\address[rvt]{School of Management and Economics, Beijing Institute of Technology, Beijing 100081, China}
\address[rvt1]{Center for Energy and Environmental Policy Research, Beijing Institute of Technology, Beijing 100081, China}
\address[cot]{Sustainable Development Research Institute for Economy and Society of Beijing, Beijing Institute of Technology, Beijing 100081, China}
\address[els]{School of Electrical Engineering, Northeast Electric Power University, Jilin 132012, China}
\address[focal]{Energy Systems Division, Argonne National Laboratory, Lemont, IL 60439, USA}

\begin{abstract}
Energy efficiency technologies (EETs) are crucial for saving energy and reducing carbon dioxide emissions. However, the diffusion of EETs in small and medium-sized enterprises is rather slow. Literature shows the interactions between innovation adopters and potential adopters have significant impacts on innovation diffusion. Enterprises lack the motivation to share information, and EETs usually lack observability, which suppress the inter-firm influence. Therefore, an information platform, together with proper policies encouraging or forcing enterprises to disclose EET-related information, should help harness inter-firm influence to accelerate EETs' diffusion. To explore whether and how such an information platform affects EETs' diffusion in small and medium-sized enterprises, this study builds an agent-based model to mimic EET diffusion processes. Based on a series of controlled numerical experiments, some counter-intuitive phenomena are discovered and explained. The results show that the information platform is a double-edged sword that notably accelerates EETs' diffusion by approximately 47\% but may also boost negative information to diffuse even faster and delay massive adoption of EETs. Increasing network density and the intensity of inter-firm influence are effective to speed EET diffusion, but their impacts diminish drastically after reaching some critical values (0.05 and 0.15 respectively) and eventually harm the stability of the system. Hence, the findings implicate that EET suppliers should carefully launch their promising but immature products; policies that can reduce the perceived risk by enterprises and the effort to maintain an informative rather than judgmental information platform can prominently mitigate the negative side effects brought by high fluidity of information.
\end{abstract}


\end{frontmatter}

\section{Introduction}

Under the stress of energy security and threat of climate change, it has become necessary to transform the world economy into a low-carbon economy. Countries to the United Nations Framework Convention on Climate Change (UNFCCC) pledged to prevent climate change and limit global warming to well below 2ºC by 2035, which indicated there still would be 13.7 billion tons of carbon dioxide (${\rm CO_2}$) worldwide, about 60\% above the level needed to remain the goal \cite{IEA2016}. To cut down ${\rm CO_2}$ emissions, improving energy efficiency is essential \cite{yang2017does}. International Energy Agency (IEA) predicts that about half of the accumulative emissions mitigation for the 2 ºC target can be achieved via the improvement of energy efficiency \cite{IEA2011}. 

Energy efficiency technologies (EETs) have been admitted as promising means to mitigate energy consumption and related ${\rm CO_2}$ emissions. Unfortunately, the diffusion of EETs is still unsatisfactory, especially among small and medium-sized enterprises (SMEs), despite the fact that they are cost-effective \cite{Fleiter2012}. The main reason is that SMEs usually allocate energy-efficiency projects with low priority and devote inadequate resources to energy management, thus manifest weak willingness to adopt EETs \cite{Cagno2013}. Considering that SMEs are quite numerous with extensive economic activities and represent the backbone of regional competitiveness \cite{Kostka2013}, effectively steering or accelerating EET adoption in SMEs is critical from the perspectives of both competitiveness and global emission-reduction targets.

Existing literature on EET adoption shows great interest in the identification and categorization of the barriers that hinder the adoption of EETs in enterprises. 
According to Cagno et al. \cite{cagno2013novel}, such barriers can be classified into mainly seven categories (for more details see \cite{Trianni2016}), among which the information related issues are a critical type of barriers frequently mentioned in literature \citep[e.g.,][]{Schleich2009,Kostka2013,nagesha2006barriers}. A recent paper summarizes that the information barriers include the cost of obtaining EET-related information \cite{rohdin2006barriers}, lack of information about energy consumption patterns \cite{schleich2008beyond}, lack of information about EET opportunities \cite{Trianni2012}, poor information for energy efficiency decisions \cite{Trianni2012}, and information issues on energy contracts \cite{trianni2013empirical}. Although these studies provide rich empirical knowledge for identifying the factors influencing enterprises’ decision on EET adoption from microscopic perspective, they are basically static without considering how enterprises change their decisions under different conditions. 

Another branch of research is from the macroscopic perspective and focuses on the aggregate patterns emerging from individual enterprises' adoption behaviors. A well-known model describing these aggregate patterns of technology adoption is the Bass diffusion model \cite{BASSFRANK1969}. The Bass diffusion model has been widely applied to the prediction and evaluation of innovation diffusion \citep[e.g.,][]{massiani2015choice,ntwoku2017ict,wang2017nls}. The original Bass diffusion model is a variant of epidemic models, which mimic the infection processes in populations \cite{BASSFRANK1969}. The Bass diffusion model is expressed in a differential equation and treats a large group of potential technology adopters as a dynamic system. Instead of following the currently prevalent research paradigm of EET adoption, the Bass diffusion model only distinguishes two types of factors that dominate the evolution of a diffusion system. The two dominant factors are external influence (e.g., advertisement and mass media) and internal influence (e.g., learning, imitation and social pressure) imposed on potential technology/innovation adopters \cite{BASSFRANK1969,Elliott2004,Kumar2016}. 

As to enterprises, internal influence, which can be interpreted as the inter-firm influence driven by competitive pressure \cite{Olupot2014}, is constrained by the inadequacy of information flows among enterprises. Indeed, market competition is a critical factor that hampers the information flows because enterprises lack the motivation to share valuable information with competitors. Moreover, EETs normally lack observability, which impedes enterprises' mutual observation and imitation. As pointed out in Rogers innovation diffusion theory \cite{Rogers1962}, the observability of an innovation is one of the five most important factors (another four factors are relative advantages, complexity, triability and compatibility) that influence an individual' adoption decision. The lack of inter-firm influence might be a crucial reason to explain why the existing literature about EET or EEM adoption intensively focuses on the organizational level instead of treating a cluster of enterprises as an organic system. 

On the one hand, previous studies provide solid empirical evidences for the situation that the lack of information is one of the most important barriers to EET adoption in enterprises; on the other hand, Rogers innovation diffusion theory as well as the Bass diffusion model place great emphasis on the importance of information flows. Therefore, it can be inferred that leveraging the inter-firm information exchange could dramatically accelerate the diffusion of EETs among SMEs. Practically, it would also  be feasible to make appropriate policies and regulations to activate the information disclosure behavior of enterprises because previous studies have confirmed that many enterprises' reported EETs were not close to organizations' core business \cite{Fleiter2013,harris2000}, and thus the disclosure of EET-related information would not harm their competitiveness. Motivated by the conflict between informational inadequacy and need, this paper aims to explore an innovative measure to facilitate EET diffusion in SMEs by answering the “ what-if ” question --- what if there exists an information platform where SMEs can share their information of EETs? Specifically, this paper targets the three questions below: 
\begin{enumerate}[label=(\roman*)]
	\item \textit{Can this information platform accelerate EET diffusion?}
	\item \textit{Will negative reports posted on the platform hamper EET diffusion?}
	\item \textit{If the answer is positive, how should negative information be handled by system designers or policymakers?}
\end{enumerate}

This paper uses agent-based modeling (ABM) to explore these questions because ABM is a bottom-up methodology and particularly effective in exploring what-if questions.  A series of controlled numerical experiments are conducted to progressively uncover whether and how an information platform impacts the diffusion of EETs in SMEs. Considering existing literature on EET diffusion is comparatively scarce, this paper proposes a novel measure to facilitating EET diffusion; also, the impacts and underlying mechanism of this measure are investigated. Three counter-intuitive effects are uncovered -- rejecter plateau", "saturation effect", and "curling effect" -- which deepen the understanding of the dynamics of EET diffusion systems. Practically, this paper sheds light on both the negative and positive impacts caused by inter-firm influence on EET diffusion, which could help policymakers foresee some probable consequences and proactively make effective policies to maintain a robust information platform that accelerates EET diffusion. Moreover, EET practitioners could extract useful information from the findings and ex-ante adjust product strategies to take advantage of the merits of adequate information flows but avoid the risks brought by its downside.

The rest of the paper is organized as follows. Relevant literature is reviewed in Section 2. Section 3, i.e. Methodology section, contains two subsections: the first subsection expounds the conceptual foundation of our model; the second presents the flowchart of the model and elaborates the model's dynamics. Section 4 calibrates and validates the model. Based on the calibration and validation, four experiments are discussed in this section, which deal with the different but connected questions mentioned above. The first and the most basic question is whether an information platform affects the diffusion pattern of EETs. We find the information platform indeed accelerates the diffusion process, but also makes the whole system sensitive to negative information and may cause a phenomenon we define as “rejecter plateau”, which inspires Experiment 2 to examine how initial performance of EETs affects the diffusion when the information platform exists. To delve deeper, Experiment 3 investigates how the number of initial disappointed adopters and risk attitude of the agents impact the diffusion pattern of EETs. Experiment 4 is conducted based on a wide parameter range and comprehensively examines the impact of network density and the intensity of one enterprise's influence on another, which deepens the understanding of inter-firm influence. Section 5 discusses the results of experiments and gives suggestions for policymakers and practitioners. Section 6 concludes this paper by summarizing the answers to the three research questions raised above and the policy as well as management suggestions discussed in Section 5.

\section{Literature review}

EET adoption is a research hotspot in recent years. Plenty of literature uses empirical methods and survey data to comprehensively analyze critical barriers that prevent enterprises from adopting EETs. Fleiter et al. \cite{Fleiter2012} analyzed barriers to the adoption of energy efficiency measures (EEMs) in SMEs based on the data from German energy audit program and found that high investment costs and lack of capital are crucial factors impeding the adoption of EEMs. Hassan et al. \cite{hassan2017barriers} investigated various barriers to the implementation of industrial EEMs using data from 192 SMEs in Pakistan and revealed that organizational sizes, qualification level of managers and technical issues have significant impacts on EEMs adoption by SMEs.
Manrique et al. \cite{manrique2018analysis} analyzed the barriers that prevent the implementation of EEMs in the ceramic sector in Colombia and found that hidden costs, followed by corporate values are the greatest barriers in this sector. Based on the rich empirical studies on EET adoption, Cagno et al. \cite{cagno2013novel} categorized the barriers into seven types: economic, information-related, organizational, behavioral, competence-related, technology-related and awareness.

It is noteworthy that the informational issues are a critical type of barriers frequently mentioned in previous literature. Nagesha and Balachandra \cite{nagesha2006barriers} argued that information barriers are crucial impediments to energy efficiency in small industry clusters based on the perceptions and value judgment of entrepreneurs. Basing on the empirical evidence from Italian primary metal manufacturing enterprises, Trianni et al. \cite{trianni2013innovation} found information barriers are main issues affecting the adoption of EETs by SMEs. Kostka et al. \cite{Kostka2013} conducted a survey involving 480 SMEs in China and demonstrated that informational barriers are the core hindrance keeping enterprises from energy-efficiency improvement. Rohdin and Thollander \cite{rohdin2006barriers} pointed out that the cost of obtaining information about the energy consumption of purchased equipment is a large barrier to energy efficiency in Swedish non-energy intensive sector. Schleich and Gruber \cite{schleich2008beyond} found that the lack of information on energy consumption pattern is the main barrier to EET adoption in commercial and service sector in Germany. Trianni and Cagno \cite{Trianni2012}
demonstrated that scarce information regarding energy efficiency opportunities and decisions is  major barrier that limits the extensive adoption of EETs by non-energy intensive manufacturing SMEs. Trianni et al. \cite{trianni2013empirical} also confirmed information issues on energy contracts are major obstacles to adopting EETs in Italian manufacturing SMEs.

From these empirical studies, it is not surprising that the speed of EET diffusion is still unsatisfactory. This is because the theory of innovaiton diffusion \cite{Rogers1962} and the Bass diffusion model \cite{BASSFRANK1969} indicate that information is a critical factor that significantly influences the diffusion among a large group of individuals. Innovation diffusion is an aggregate pattern emerging from each enterprise's adoption of new technologies. The process of innovation diffusion resembles the transmission of infectious diseases among a large group of individuals. The Bass diffusion model is well-known for its good performance in describing real-world innovation diffusion patterns, which is originally inspired by epidemic models \cite{BASSFRANK1969}. The Bass diffusion model treats a large group of potential technology adopters as a dynamic system whose evolution is essentially driven by external influence (e.g., advertisement and mass media) and internal influence (e.g., learning, imitation and social pressure) of information \cite{BASSFRANK1969}. Bass assumed that the number of new adopters at time $t$ depend linearly on the external and internal influence, which can be expressed as the differential equation Eq. (\ref{eq:bass_model}). 
\begin{equation}
y(t)=(p+q\times Y(t)/N)\times(N-Y(t))\label{eq:bass_model}
\end{equation}
where $y(t)$ represents the new adopters at time $t$; constant $p$ represents the external influence from mass media and reflects the importance of innovators who adopt innovation spontaneously; constant $q$ represents the internal influence from, e.g., learning, imitation, social pressure, and network effects \cite{Kiesling2012,ganjeizadeh2017applying}; constant $N$ is the total number of potential adopters for new products on the market; $Y(t)$ is the number of previous adopters. Eq. (\ref{eq:bass_model}) implies that the importance of $p$ is greatest in the beginning of the diffusion process (because there are many non-adopters) but monotonically recedes with time; while $q$ exerts increasing influence on potential adopters as the number of cumulative adopters grows. Many previous studies interpret the $q$ as word-of-mouth effect \cite{Zhang2011,ramirez2019forecasting}. As to enterprises, the $q$ could be interpreted as the inter-firm influence driven by competitive pressure \cite{Olupot2014,Oliveira2010}, which is constrained by the inadequacy of information flows among enterprises. 

The Bass diffusion model mathematically demonstrates the importance of information propagation to the diffusion of new technologies. Numerous studies have applied this model (or its variants) to the exploration, prediction, and evaluation of a variety of energy-related technologies \citep[e.g.,][]{dalla2014diffusion,she2019analysis,reddy2018economic}. However, sparse literature regarding EET adoption/diffusion focuses on the impact of inter-firm influence.  A pertinent research is \cite{arens2014diffusion}, which uses the empirical data of several typical EETs to explore how the diffusion rates impact the trend of energy intensity development in the German steel industry but does not place emphasis on the inherent mechanism of EET diffusion. More relevant literature regarding the diffusion of energy-related technologies can be found in the domain of renewable energy \citep[e.g.,][]{eleftheriadis2015identifying,zeng2018disruptive,horbach2018energy}. Nevertheless, existing studies emphasizing mutual influence between individuals mostly focus on the technology diffusion in large groups of consumers \cite{palmer2015modeling,hyysalo2017diffusion,stavrakas2019agent}, who are generous to share their experience of novel things. Different from consumers, enterprises lack the motivation to share valuable information with competitors, which hampers inter-firm information flows. Another crucial reason to explain the slow diffusion speed of EETs as well as the sparsity of literature on this topic might be that EETs normally lack observability. Rogers \cite{Rogers1962} pointed out the observability is one of the most important driving forces for potential adopters to accept innovations through observations and imitations. The gap between the importance of information to EET diffusion and the current inadequacy of inter-firm information flows motivates this study to investigate the impact of an information platform on EET diffusion.

Although the Bass diffusion model provides sufficient mathematical simplicity and clear categorization of driving forces of technology diffusion, it is not suitable for answering the what-if questions \cite{Byrka2016}.  Actually, the Bass diffusion model and its variants are differential-equation-based models, which treat the diffusion system as a whole but lack the ability of modeling heterogeneous decision-making processes of individuals. Researchers notice aggregate diffusion models' shortage of explanatory power \cite{Kiesling2012}. As Hohnisch et al. \cite{hohnisch2008percolation} pointed out, the Bass model is phenomenological and does not reflect the underlying mechanism of innovation diffusion.  A typical innovation diffusion trajectory is an S-shaped curve, which reflects the diffusion rate of a technology over time. An S-shaped diffusion pattern is the aggregate result of the interactions between the potential adopters, who are essentially decision-makers. That is to say the regularities of technology diffusion at system-level result from the activities at individual level. To demystify the underlying mechanism of EET diffusion systems, it is necessary to bridge the system level and individual level instead of dealing with the two levels separately. This thought leads to the necessity of applying agent-based modeling, which  models the decision and adaption behaviors of each individual at microscopic level but let the regularities emerge spontaneously at system/macroscopic level. 

ABM is particularly effective in exploring dynamic systems that consist of numerous adaptive and interactive individuals. This property of ABM widely attracts researchers from various areas. Many researchers, interested in innovation and technology diffusion also unexceptionally, adopt ABM as a convenient and powerful tool to explore theoretical insights, predict diffusion rates, and evaluate policy impacts. Goldenberg et al. \cite{Goldenberg2007} employed an agent-based model to analyze the effect of negative word-of-mouth on firms' profits simultaneously from individual and network levels. Rand and Rust \cite{Rand2011} built an agent-based model to fit the Bass diffusion model with empirical data and proposed a framework for rigorously building agent-based models. Silvia and Krause \cite{silvia2016assessing} built an agent-based model to simulate the individual adoption decisions of plug-in electric vehicles under four policy scenarios aimed at promoting electric vehicle adoption. Zeng et al. \cite{zeng2018disruptive} used ABM to simulate the competition dynamics between renewable energy technologies and conventional energy technologies. Karslen et al. \cite{karslen2019agent} developed an agent-based model to explore how split incentives barriers affect the diffusion of Flettner rotor technology in shipping.  

Rand et al. \cite{Rand2011} listed 6 criteria to determine when agent-based models are more suitable than aggregate models. The most important (also a sufficient one) criterion is that the target system consists of many adaptive agents. Since enterprises react correspondingly to the variations of environment and influence from other entities, it is suitable to use ABM to explore the impact of an information platform on the EET diffusion among enterprises.

\section{Methodology}
The Bass model provides a clear-cut foundation of classifying the driving forces as well as a stylized pattern of EETs' diffusion. However, its clarity is accompanied with oversimplification, which consequently veils the concrete mechanism of innovation diffusion. We use the taxonomy of the Bass diffusion model as a starting point but break down the external influence ($p$) and internal influence ($q$) into more detailed factors to incorporate enterprises' decision-making processes and interactions.

\begin{figure}[H]
	\centering
	\includegraphics[width=0.4\linewidth]{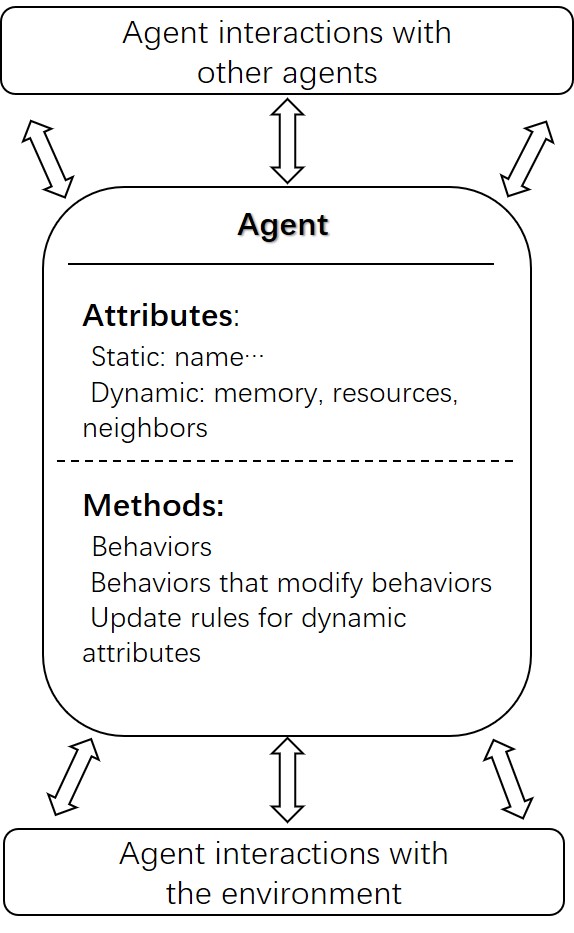}
	\caption{A typical agent \protect{\cite{Macal2011Introductory}}}
	\label{fig:typicalagent}
\end{figure}

A typical agent-based model is mainly constituted of three elements (see Figure \ref{fig:typicalagent}): 1) a cluster of agents with specific attributes and behaviors; 2) a cluster of agent relationships and methods of interactions which describe how and with whom agents interact; and 3) the environment where the agents behave \cite{Muehleisen2016}. To map the three elements to our specific research topic, i.e., the EET diffusion among SMEs, we introduce the framework of technological, organizational, and environmental context (TOE) proposed by Tornatzky et al. \cite{Tornatzky1990} to build the model's conceptual foundation. 

The TOE framework was developed to analyze information technology adoptions at firm level \cite{oliveira2011literature}. It is unnecessary to apply the TOE framework specified with exactly the same details as its original version. Relevant literature shows the framework has been fleshed out with different details in various scenarios \cite{Oliveira2010,Oliveira2009,Lee2009}. Herein, the TOE framework works as a reference frame for the model to incorporate necessary components to build an organizational adoption model.  With the assistance of the TOE framework, one can also easily locate what elements in this framework are not implemented in the current model. As to the environment context, we use a random network comprised of a fixed quantity of enterprises to simulate the market structure. Regarding the technological context, we condense the technological characteristics of EETs proposed by Fleiter et al. \cite{Fleiter2012a} into two main categories, i.e. static type and dynamic type represented by a two-dimensional vector in our model. Accordingly, the organizational context in terms of EETs can also be represented by a vector with two dimensions which respectively indicate the requirements of the static technological characteristics and demands for the dynamic technological characteristics by SMEs.

\begin{figure}[H]
	\centering
	\includegraphics[scale=0.5]{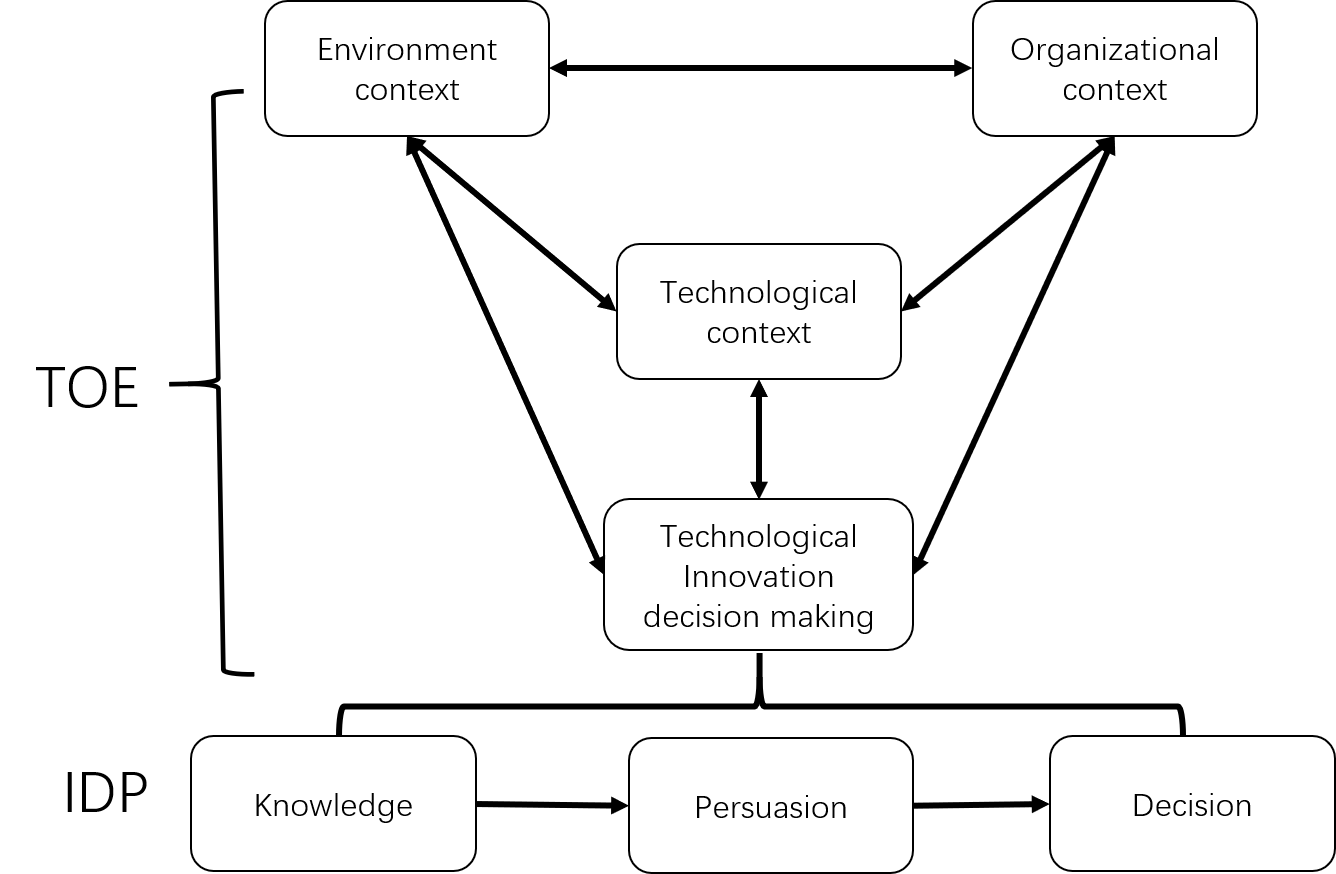}
	\caption{The TOE and IDP framework}
	\label{fig:TOE_IDP}
\end{figure}

In order to clarify the dynamics of how these components work together along the time line, the TOE framework is further mapped to the innovation decision process (IDP) proposed by Rogers \cite{Rogers1962} (see Figure \ref{fig:TOE_IDP}). The original innovation decision process consists of five stages: knowledge, persuasion, decision, implementation, and confirmation, among which the first three stages determine whether a decision maker adopts the innovation or not. Thus, it is sufficient to focus on the first three stages for this research. In the knowledge stage, potential adopters become aware of an innovation under the influence of information from massive media and other enterprises. In our model, the external influence and inter-firm influence (via random network) are effective in this stage. Once an enterprise becomes aware of the innovation, the process proceeds to the persuasion stage, and the enterprise begins to investigate whether this technology matches its requirements and demands. Therefore, the technological and organizational characteristics mentioned in the TOE framework are mapped to this stage. Finally, the higher the technology-organization match degree is, the more likely an enterprise tends to adopt the technology, which rightly fits into the decision stage of the innovation decision process. The details about model building are described as follows.

\subsection{Model's components}
\subsubsection{The technological context}
EETs with different characteristics have distinct diffusion processes \cite{Fleiter2012a,Trianni2014}. Fleiter et al. \cite{Fleiter2012a} built a framework to classify EETs with different characteristics, which consists of three criteria, i.e., relative advantage, technical context, and information context. The three criteria are further divided into 12 detailed characteristics. Although, these characteristics are critical to defining an EET, existing literature does not give an equation to precisely describe the impact of each characteristic on enterprises' decisions. Under this situation, literally copying all these factors into the model cannot enhance its explanatory power. To incorporate the characteristics of EETs but avoid inducing unnecessary complexity of model building, these characteristics are re-categorized into two types: dynamic and static (see Table \ref{table:Fleiter_scheme}). The internal rate of return, payback period, and initial expenditure etc., are highly changeable as the EET becomes mature; while distance to core process, type of modification, and scope of impact etc., are qualitatively constrained by the technological paradigm an EET belongs to, which means these characteristics are comparatively stable \cite{dosi1982technological}. For example, light-emitting diodes (LEDs) are a typical energy-efficient technology. The U.S. Department of Energy estimated that widespread use of LEDs could save about 348 TWh (compared to no LED use) of electricity in the US by 2027 \cite{ledestimate}. According to \cite{ledreport}, the price of LED bulbs dropped rapidly from 37\$/klm in 2012 to 9 \$/klm in 2016. Opposite to the decreasing trend of prices, the U.S. Energy Information Administration predicted LED bulb efficiency in 2030 would be three times higher than that in 2010 \cite{ledinform}.  Higher performance together with lower price of LED technology significantly impact its internal rate of return, payback period, initial expenditure etc over time. However, illuminations are important to factories that involve intensive cutting, drilling, and punching processes because well illuminated workplace can effectively enhance productivity and production safety. Therefore, these factories may give high priority to illumination investment. The technological development of LEDs (as an illumination technology)does not fundamentally change LEDs’ distance to these factories' core business and scope of impact etc. Thus, these characteristics can be regarded as static.  

The static-dynamic taxonomy is not intended to precisely distinguish every characteristic of an EET, but it follows the thought of Dosi technological paradigm and technological trajectory \cite{dosi1982technological}, which reflect a technology's static and dynamic aspects. With this categorization, a two-dimensional vector $\vec{T}$ is introduced to represent the two distinct types of characteristics of EETs:
\begin{equation}
\vec{T}=[T_s,T_d]
\label{eq:tech_vector}
\end{equation}                                                
where $T_s$ represents the static technological characteristics; $T_d$ stands for the dynamic characteristics, which is calculated as
\begin{equation}
T_d = a(E_d - b)\times Y(t)/N+b
\label{eq:tech}
\end{equation}                                                                
where $E_d$ is the demands of an enterprise for dynamic characteristics; $b$ denotes the initial performance of EETs when they firstly enter the market; constant $a$ is a coefficient adjusting the speed of technology progress; $Y(t)$ denotes cumulative adopters of EETs at year $t$; $N$ denotes the total number of potential adopters for an EET in a market. This equation reflects the dynamic feature of technology development: the more products are sold, the more revenue a producer obtains, and thus more research and development (R\&D) resource can be devoted to improving products. When $T_d = E_d$ , then $a = N/Y(t)$, which could offer a more intuitive explanation of $a$. $a$ means the proportion of $N$ to $Y(t)$ when technologies' dynamic characteristics can completely satisfy enterprises' dynamic demands. For example, $a=1$ means a technology can completely satisfy enterprises' dynamic demands when all of the considered enterprises adopt it; $a=2$ means a technology needs a half of all the considered enterprises to adopt it to become fully mature. Since many successful technologies tend to oversupply their performance \cite{christensen2013innovator}, we expect $a>1$, which means technologies become completely mature before everyone becomes an adopter in this model.

\subsubsection{The organizational context} 
Organizational characteristics have distinct impacts on the inclination towards innovation adoptions. The classification scheme for energy efficiency measures proposed by Fleiter et al. \cite{Fleiter2012a} provides a perspective for identifying not only EET-related technological characteristics but also for capturing organizational characteristics needed for the analysis of EET adoption (see Table \ref{table:Fleiter_scheme}). For instance, “distance to core process”, “type of modification”, “scope of impact” etc place emphasis on both technologies and enterprises. These characteristics do not change very much with the progress of technologies. Other characteristics, such as “internal rate of return”, “initial expenditure”, and “payback period” etc are changeable with the development of EETs, which can dynamically change the risk and profit perceived by enterprises and then affect their adoption decisions. Similar to the Eq. (\ref{eq:tech_vector}) used to describe technological characteristics, we use a two-dimensional vector to represent the characteristics of enterprises.
\begin{equation}
\vec{E}=[E_s,E_d]
\end{equation}
where $E_s$ denotes the static EET-related characteristics of enterprises, and $E_d$ denotes enterprises' demands for dynamic technological characteristics of EETs.
\begin{table}[ht]
	\caption{Classification scheme for energy efficiency measures \protect\cite{Fleiter2012a}} % title of Table
	\centering % used for centering table
	\begin{tabular}{cccc} % centered columns (4 columns)
		\hline%\hline %inserts double horizontal lines
		& Characteristics & Static & Dynamic \\ [0.5ex] % inserts table
		%heading
		\hline % inserts single horizontal line
		\textbf{Relative advantage} & Internal rate of return &  &\checkmark  \\ % inserting body of the table
		& Payback period          &  &\checkmark \\
		& Initial expenditure     &  &\checkmark\\
		& Non-energy benefits 	  & \checkmark & \\ [1ex] % [1ex] adds vertical space
		\textbf{Technical context} & Distance to core process & \checkmark &  \\
		& Type of modification & \checkmark &  \\
		& Scope of impact & \checkmark &  \\
		& Lifetime & &  \checkmark \\				[1ex] % [1ex] adds vertical space
		\textbf{Information context} & Transaction costs &  & \checkmark \\ 
		& Knowledge for planning and implementation & \checkmark &  \\
		& Diffusion progress &  & \checkmark  \\
		& Sectoral applicability & \checkmark &  \\ [1ex] % [1ex] adds vertical space
		
		\hline %inserts single line
	\end{tabular}
	\label{table:Fleiter_scheme} 
\end{table}

\subsubsection{The environmental context} 
Environmental context mainly includes industry characteristics and market structure, coping with the government, and technology support infrastructure. It is noteworthy that the environment herein refers to the external conditions of (potential) EET adopters instead of EET suppliers. EET adopters come from a diversity of industries, and their characteristics may differ significantly. It is infeasible to implement all the details of environmental context in the model. Moreover, there does not exist an admitted representative for diverse industry characteristics.  Following the KISS (keep-it-simple-stupid) principle of complex adaptive system modeling \cite{axelrod1997complexity}, this study focuses on a universal characteristic of (potential) EET adopters: the inadequacy of inter-firm information flows due to EETs’ lack of observability and the competitive relationships between enterprises. To mimic this characteristic, enterprises are placed in a sparse complex network, and the probabilities of inter-firm interactions are assigned with small values. Considering the agents are enterprises of small and medium sizes, a random network is suitable for modeling their market positions because nodes are statistically equivalent in random networks, which corresponds to the fact that each SME has no dominant market power over others (different from scale-free networks). Government policies have great impact on the risk attitudes of market participants and are modeled with a global variable denoted as “$n$” that adjusts the risk preferences of enterprises. For simplicity, technology support infrastructure is assumed to be able to completely satisfy the EET adopters. This assumption could be relaxed in the future research. 

Regarding the network in which enterprises interact with their counterparts (e.g., competitors, suppliers and consumers), we use “$q_s$”, which is similar but not identical to the Bass $q$, to denote the probability that an individual enterprise is successfully influenced by its linked enterprises. Assuming that enterprises should post their experience of using a specific EET honestly on the platform, it is natural to consider simultaneously the effect of negative and positive comments. Positive comments can propel the diffusion of an innovation, while negative comments can do the contrary. However, negative comments are useful for enterprises to avoid inferior technologies and further to save resources. To be more realistic, Goldenberg et al. \cite{Goldenberg2007} and Wang et al. \cite{WANG2018} proposed an algorithm of innovation diffusion, which takes into account the influence of negative word-of-mouth. Therefore, every individual in the system is exposed to the influence of positive and negative information. The probability that enterprise $i$ is under the influence of positive information is determined as Eq. (\ref{eq:positive}).
\begin{equation}
pr^+_i =1-(1-p_s)(1-q_s)^{As_i (t)}
\label{eq:positive}
\end{equation}
where the $p_s$ indicates probability of being under the external influence including advertisement, the incentive measures and pressure from government policy, which should always be positive; $As_i(t)$ is the number of satisfied adopters linked to enterprise $i$ at time $t$. Contrarily, the probability that enterprise $i$ is under the influence of negative information is calculated as Eq. (\ref{eq:negative}).
\begin{equation}
pr^-_i =1-(1-nq_s)^{Ad_i (t)}
\label{eq:negative}
\end{equation}
where $Ad_i (t)$ is the number of disappointed adopters linked to enterprise $i$ at time $t$; as mentioned above, $nq_s$ in this equation means that negative information is as $n$ times more influential as the positive. According to these two equations, the inter-firm influence on an specific enterprise is determined by two variables, i.e., $q_s$ and the numbers of influential neighbors, i.e., $As_i$ or $Ad_i$. The $q_s$ quantifies the intensity of influence that one enterprise exerts on one another; while the numbers of influential neighbors indicates how many such enterprise would influence one enterprise. For clarity, we label the $q_s$ as “interactive coefficient” to distinguish the $q$ in Bass model. Hence, it is clear that the inter-firm influence in our model is comprised of two elements: the “interactive coefficient” and the number of influential neighbors.

Enterprise $i$ may be affected by positive, negative, and both positive and negative information, or neither. Having modified the equations of Goldenberg et al. model \cite{Goldenberg2007} for this present study, the probabilities of being influenced by different types of information are calculated using Eq. (\ref{eq:absorb}) to (\ref{eq:non-adoption}). $(1-pr_i ^- )pr_i ^+$ is the probability of being affected by only positive information; while  $(1-pr_i^+)pr_i^-$ is the probability of being affected only by negative information. The probability of being affected by both positive and negative information is $pr_i^+ pr_i^-$, among which, a certain proportion of $u_i$ ($u_i=pr_i^+/(pr_i^+ + pr_i^-)$) equals the probability that an enterprise decides to give a further consideration of EETs, with probability of $1-u_i$ to reject the technology directly. 
\begin{equation}
Pa_i(t)=(1-pr_i^-)pr_i^+ + u_ipr_i^+pr_i^-
\label{eq:absorb}
\end{equation}
\begin{equation}
Pr_i(t)=(1-pr_i^+)pr_i^- + (1-u_i)pr_i^+pr_i^-
\label{eq:reject}
\end{equation}
\begin{equation}
Pn_i(t)=(1-pr_i^+)(1-pr_i^-)
\label{eq:non-adoption}
\end{equation}

Note, it is assumed that enterprises rejecting EETs are not similar to human in the sense that enterprises would spread negative information of a certain product merely because they are influenced by others’ negative word-of-mouth. Therefore, in the present model, only the disappointed adopters who really have bad experience of EETs can disseminate negative information.

\subsubsection{The decision-making process of SMEs}
\begin{figure}[H]
	\centering
	\includegraphics[scale=0.27]{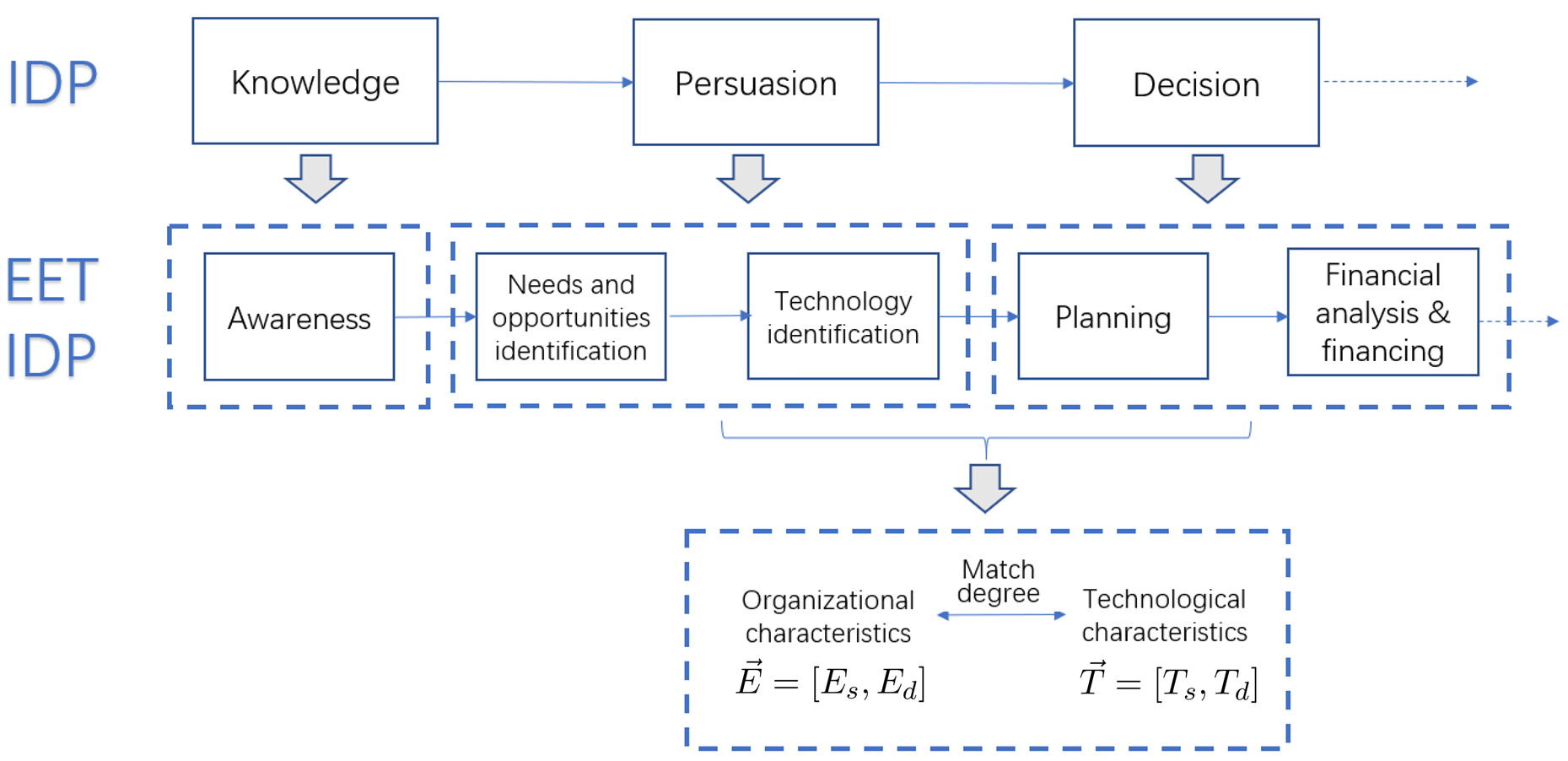}
	\caption{From IDP and EET IDP to technology-organization match}
	\label{fig:EET_IDP}
\end{figure}

An EET decision-making process has been derived from the general innovation decision-making framework of Rogers \cite{Trianni2016,Rogers1962}. As shown in Figure \ref{fig:EET_IDP}, the first step of EET decision-making process is awareness. As mentioned above, an enterprise’s awareness of EETs is determined by the external influence and the internal influence, and calculated as Eq. (\ref{eq:absorb}). If the enterprise becomes aware of EETs, the process proceeds to “needs and opportunities identification” and “technology identification” stages. The “needs and opportunities identification” stage means the organization identifies its needs according to its characteristics and tries to look for proper EETs, which leads to the “technology identification” stage when the organization intends to find the EETs that match best the characteristics of the organization and its demands for technological maturity. In this paper, we use the similarity between the technology vector $\vec{T}=[Ts,Td]$ and enterprise vector $\vec{E}=[Es,Ed]$ to measure the technology-organization matching degree:
\begin{equation}
Pm_i(t)=(|\vec{T}|\times\cos\theta)/|\vec{E}|
\end{equation} 
where $\theta$ is the angle between vectors. 
This two stages together correspond to the persuasion module of Rogers’ model. The decision stage determines whether an enterprise adopts EETs: higher match degree means higher probability to adopt, and vice versa. 

The planning and financial factors related to decision-making are of great interest and importance, but not the focus of this paper, thus are not considered. This should not impair the effectiveness of this model because the Bass model considers nothing about decision-making \cite{BASSFRANK1969,BASSFARANK1994} but still works well; and Rand \cite{Rand2011} built an agent-based model without considering decision module had been well calibrated and validated. 
\begin{figure}[H]
	\includegraphics[width=\textwidth]{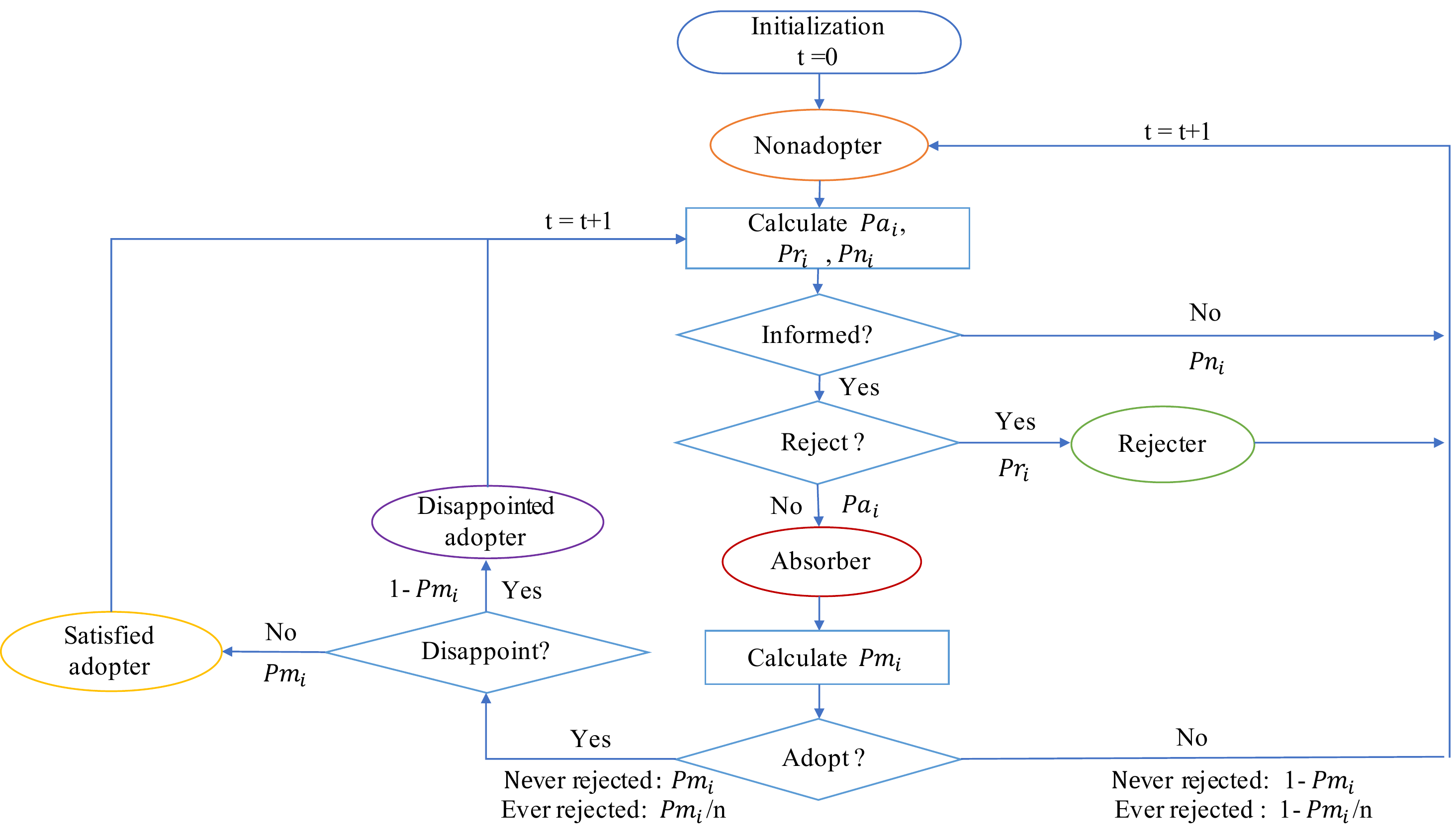}
	\caption{Flowchart of the model}
	\label{fig:flowchart}
\end{figure}  
\subsection{Model dynamics}

Figure \ref{fig:flowchart} shows how the framework is implemented. The model is built based on a random network where each agent has the same probability being linked to other agents. Each SME has five types of potential status shown in the eclipses and starts as a non-adopter. A non-adopter can be affected by the positive and negative information externally and internally. It can transform to a rejecter by a probability of $Pr_i$ or become an absorber by $Pa_i$ , who is positively informed about the technology and proceeds to calculate how much the technology matches its demands, and decides to adopt by a probability of $Pm_i$. A non-adopter can also stay intact if they are not informed, by a probability of $Pn_i$. An adopter evolves to either a satisfied adopter or a disappointed adopter by probabilities of $Pm_i$ and $(1- Pm_i)$ respectively. A satisfied adopter spreads positive information, while a disappointed adopter disseminates negative information, which affect the values of $Pa_i$, $Pr_i$ and $Pn_i$. Since the model considers EETs as a whole, rather than any specific technology or brand, the rejecters are not assumed to have any possibility to leave the market permanently,  but only be able to delay their adoption. In this sense, rejecters can become adopters. However, because the influence of negative information fosters the rejecters’ attitude, it should be more difficult for rejecters to become adopters than intact non-adopters do. Thus, “$n$”, which reflects the risk attitude of enterprises is used to modify the probability of adoption by a rejecter. As Figure \ref{fig:flowchart} shows, the probability of a rejecter to adopt EETs is $Pm_i/n$. 

\section{Experiments}
\subsection{Calibration and validation}
It is commonplace to calibrate the model with empirical data of EET diffusion in an enterprise cluster. Unfortunately, there are no such data completely suitable for the calibration. Since the Bass model has been widely admitted as an innovation diffusion model that works well to fit realistic data, we calibrate the agent-based model using a Bass model with the average $q$ value derived from 213 technologies and the  average  $p$ value of 10 energy-related technologies \cite{Sultan1990}. Under these values, the Bass model shows an S-shaped curve that reaches 95\% of market share in the $16th$ year (the years when technologies reach 95\% market share are denoted as $year_{0.95}$ in this paper). Using these data for calibration is far from perfect and it is reasonable to consider that a larger $year_{0.95}$ for EETs could be more appropriate, but they can offer a stylized S-shaped curve of technology diffusion. Given that the purpose of this paper is not predictive use but exploratory research, these data should be sufficient to provide a benchmark for avoiding aberration. Moreover, Experiment 4 has been conducted using 1000 different combinations of interactive coefficient and network density, which should be adequate for a comprehensive understanding of inter-firm influence, in spite of the lack of empirical data.

Empirical evidences can be found to narrow down the parameter space to obtain a set of reasonable calibrated parameters. We use 500 ($N$ = 500) agents to initialize the number of potential adopters (SMEs), which is supported by empirical studies conducted by Tao and Todeva \cite{Tao2006}, and Takeda et al. \cite{Takeda2008} who respectively investigated the network structure of an industrial cluster in China and Japan containing 314 and 604 enterprises. Indeed, the actual number of potential market $N$ has no qualitative effect on the pattern of EET diffusion when technology substitutions are not considered. Rand and Rust \cite{Rand2011} decreased the potential market $N$ from 40001 to 400 for facilitating the analysis of their agent-based model, and the variance measures indicated it had no qualitative influence on the results. In many previous studies, the potential market $N$ is often assumed as 1 \cite{Kiesling2012,Fleiter2013}. The values of network density ($nd$) of 0.016 \cite{Tao2006} and 0.01 \cite{Takeda2008} are found in the enterprise clusters. The 0.016 network density is actually from supply-chain-based data. Given that enterprises have little motivation to share information with others and the lack of observability of EETs, we assume that the network density should be no more than 0.016 in terms of EET diffusion. However, a wide range of network density values from 0.002 to 0.1 with intervals of 0.001 has been searched, just in case the data are not representative enough.

The Bass $q$ determines the ratio that non-adopters become adopters, whereas the $q_s$ in the agent-based model is only responsible for the knowledge stage. Therefore, to estimate a reasonable range of $q_s$ for agent-based modeling, both knowledge, decision stages and the influence of negative information should all be taken into consideration, which differ $q_s$ from the $q$ value in Bass model. Having considered these factors, we have searched a large parameter space of $q_s$ from 0.01 to 1.00  with intervals of 0.001. Similarly, $p_s$ in our agent-based model should be greater than the $p$ of Bass model because in the ABM only when an enterprise is informed and makes its decision (by a probability of $Pm_i$) to adopt, it then becomes an adopter. Thus the model has been given a range of $p_s$ values from 0.018 to 0.0445 with intervals of 0.001 for searching.

The “$n$” in our model reflects the risk attitude of an enterprise. As described above, on the one hand “$n$” let negative information become more influential than positive innovation, accordingly driving non-adopters to become rejecters; on the other hand, the “$n$” makes it more difficult to turn rejecters into adopters. According to prospect theory \cite{Kahneman1979} and previous studies \cite{Goldenberg2007,Technical1985,Hart1990}, the “$n$” is set as 2.

The EETs closer to an enterprise’s core business diffuse more slowly \cite{Fleiter2012}. Given that the combination of $p = 0.018$ and $q = 0.38$ does not show a typically slow pattern, it is reasonable to assume that averagely these energy-related technologies are not close to the enterprises’ core business, thus can easily match the characteristics of these organizations, and 40 is assigned to the first item of $\vec{T}=[T_s,T_d]$, i.e., $T_s = 40$, and the first item of $\vec{E}=[E_s,E_d]$ is set as 50, i.e., $E_s = 50$, which means these technologies are assumed to match SMEs well ($40/50=0.8$) in terms of static characteristics. A range of $a$ values from 0.1 to 3 with intervals of 0.1 has been searched, which means the technology needs at least $1/3$ of the considered enterprises adopting it to become fully mature. 

\begin{figure}[H]
	\centering
	\includegraphics[scale=0.7]{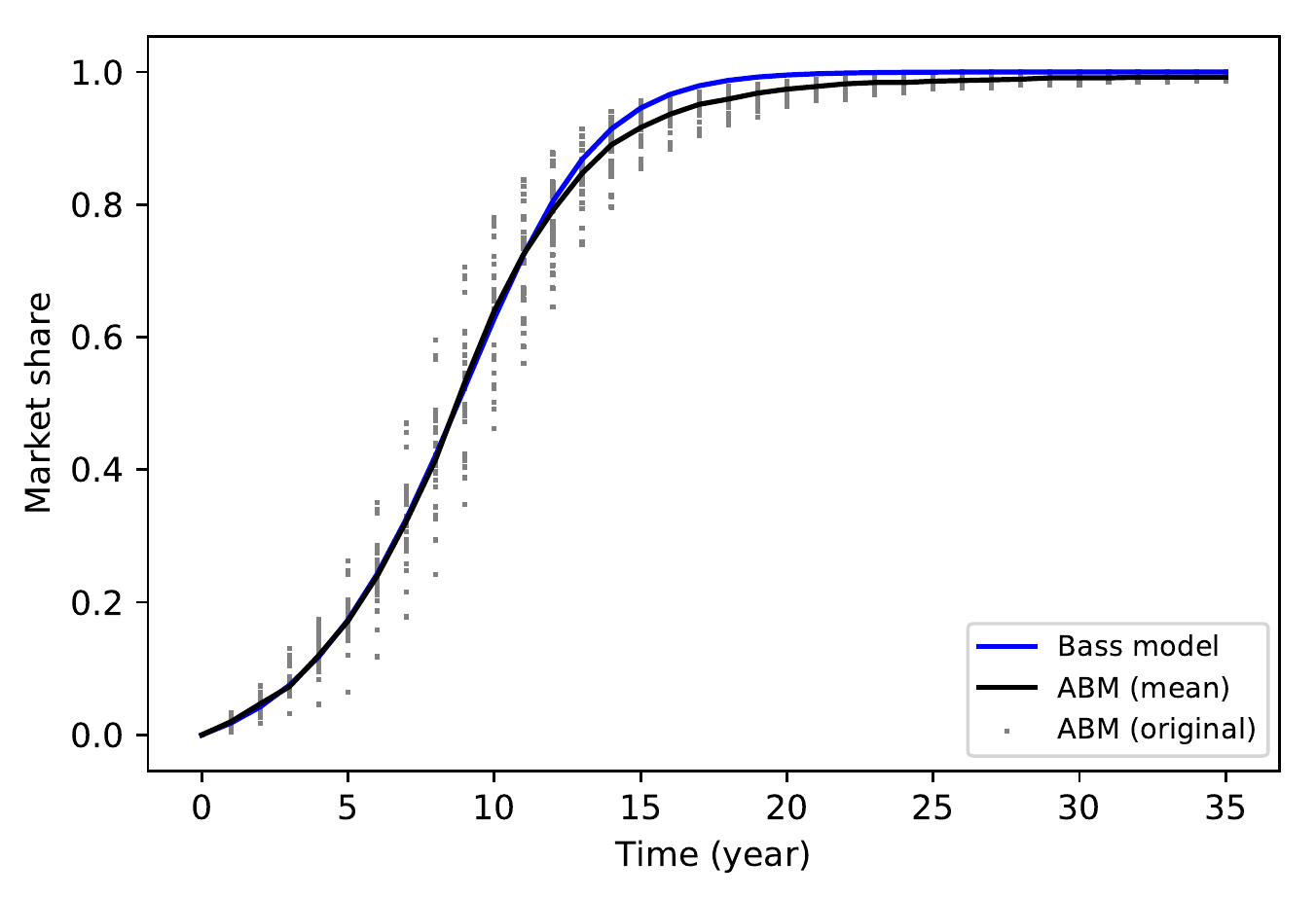}
	\caption{Fit the output of ABM with Bass model}
	\label{fig:vv}
\end{figure}

Netlogo’s Behaviorsearch \cite{Wilensky2003} is used to calibrate the model’s parameters. In order to minimize the difference between the model’s output and the target data, we use ordinary least square (OLS) as the objective function, use Behaviorsearch’ genetic algorithm to search the parameter space, and run the model for 10000 times of 8 repetitions. Consequently, the combination of $p_s=0.037$, $q_s=0.241$, $nd=0.01$, $a=2.5$, $b=30$ is obtained, with a $R^2 = 0.976$.
Hopefully, although some parameters are given a generous searching range, the calibration still results in meaningful parameter values. Figure \ref{fig:vv} shows the scatter diagram of 30 runs with the obtained parameters combination and a curve representing the median value of market share in each step. This curve is S-shaped and reflects the stylized fact that innovation diffusion starts slowly but increases dramatically when the number of adopters reaches a certain threshold (when the first turning point occurs) \cite{Karakaya2016}. After a period of time, the diffusion rate begins to slow down and tends to reach all of the non-adopters. Therefore, the model is qualitatively validated by means of cross-model validation. Moreover, the model is built based on the TOE and IDP framework which have been widely accepted by industry and academia, thus it is conceptually validated. 
ABM provides researchers with the opportunity to conduct controlled numerical experiments on complex systems. Based on the calibrated and validated model, a series of experiments are conducted to answer the question: what if there exists an information platform where SMEs can share their information of EETs? 

\begin{table}
	\caption{Parameters of the model}
	\label{tab:parameter value}
	\centering
	\begin{tabular}{llll}
		\hline
		\textbf{Parameter} & \textbf{Meaning}                                                              & \textbf{value}                                                                  \\\hline
		
		$N$       & the total number of SMEs                                             & 500                                \\ 
		$p$       & the external influence in the Bass model             & 0.018                                       \\ 
		$q$       & the internal influence in the Bass model & 0.38                                   \\ 
		$T_s$     & the static attributes of EETs                     & 40    &                                                                            \\ 
		$E_s$     & \begin{tabular}[c]{@{}l@{}} SMEs' requirements for static technological \\characteristics of EETs \end{tabular}     & 50    &                                                                            \\ 
		$E_d$     & \begin{tabular}[c]{@{}l@{}} SMEs' demands for dynamic technological \\characteristics of EETs \end{tabular}     & 80&                                                              \\                  
		$nd$      & the network density                                                  & 0.01  &                                                             \\ 
		$p_s$     & the external influence in the agent-based model                                   & 0.037 &                                                                 \\ 
		$q_s$     & the inter-firm influence in the agent-based model                                  & 0.241 &                                                                \\ 
		$a$       & the speed of technology progress                                     & 2.5   &                                                                \\ 
		$b$       & the initial performance of EETs                                      & 30    &                                                              \\ 
		$n$       & the risk preferences of SMEs                                         & 2     &                                                               \\\hline
	\end{tabular}
\end{table}
%\newpage 这条命令用于单栏排版

\subsection{Experiment 1: the impact of information disclosure on the platform}
\begin{figure}[H]
	\centering
	\includegraphics[scale=0.7]{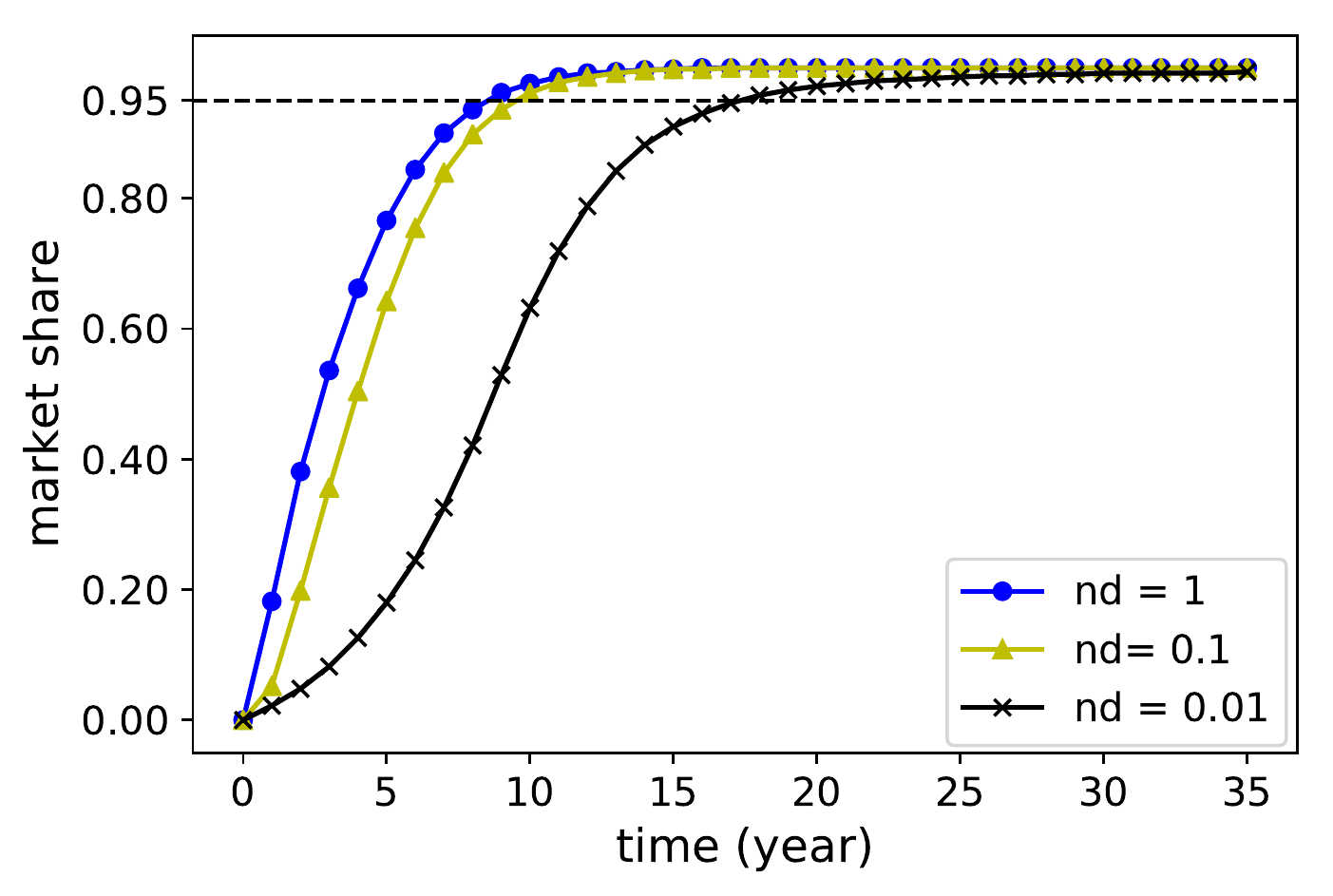}
	\caption{Experiment 1 --- The speed of EET diffusion with different network density ($nd$) } 
	\label{fig:link=1}
\end{figure}

If there exists an official information platform where SMEs are forced or encouraged to post their comments of EETs used, every enterprise in the system is able to know about the disclosed EET information directly. That is, all SMEs in the system are completely connected. Therefore, the immediate impact of the information platform is to increase network density ($nd$) to 1, and the first experiment is conducted under this condition. The model has been run for 300 times to obtain the mean value of market shares in each year when network density equals 1. As Figure \ref{fig:link=1} shows, a public information platform let every SME in the system be connected. The diffusion of EETs begins to take off as soon as the emergence of the technology. It can be witnessed that when the network density becomes 1, it only takes about 9 years to reach 95\% of market share, compared with about 17 years when the network density equals 0.01, which is a tremendous progress (reducing over 47\% diffusion period) and indicates that the disclosure of EETs on a public platform by enterprises can significantly improve the speed of diffusion. The result of $nd=0.1$ is also given in this figure to compare the impact of different network densities. It shows a diffusion curve much approximating to the result of $nd=1$. The impact of network densification is explored in detail in Experiment 4.

\begin{figure}[H]%[!tpb]
	\centering
	\captionsetup{justification=centering,margin=3cm}
	\includegraphics[scale= 0.6]{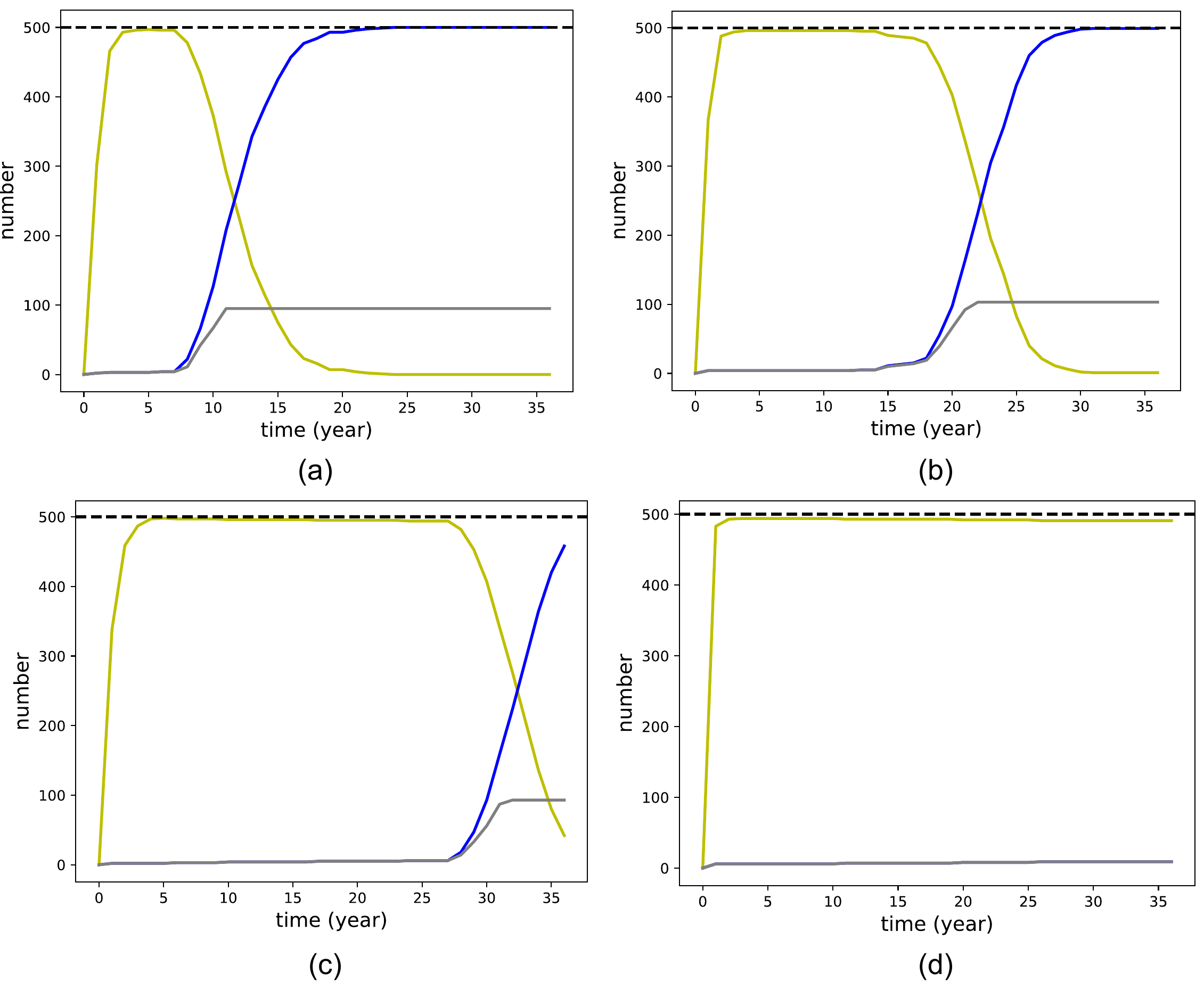}
	\caption{Experiment 1 --- Rejecter plateaus of different durations\\ blue, yellow, and gray lines represent adopters, rejecters, and disappointed adopters respectively (these results are obtained when $b=10$)}
	\label{fig:rejecters_plateau}
\end{figure}

However, each coin has two sides. The platform built to accelerate the diffusion of EETs can also drive negative information even diffuse faster than the positive does, especially under the condition of high network density. Goldenberg et al. \cite{Goldenberg2007} pointed out that negative word-of-mouth is more influential and even can determine the destiny of products. The model manifests some extreme behaviors, which can be vividly named as “rejecter plateau” phenomenon as shown in Figure \ref{fig:rejecters_plateau} (to obtain a diversity of rejecter plateaus, $b=10$ was used to generate these graphs). As the model is essentially stochastic, there exist chances that over 90\% of enterprises become rejecters typically in the first 3 years and remain rejecting EETs for a period of time, then rapidly change to be adopters once a satisfied adopter appears and broadcasts positive information. The shape of the curve is similar to a plateau with two asymmetric steep edges indicating the system is more sensitive to negative information than to positive information. As can be seen, this plateau is caused by only a few of disappointed adopters (gray lines in Figure \ref{fig:rejecters_plateau}), usually less than 10. Although it seems to be aberrant, there are some practical reasons that can explain how this phenomenon occurs.

At first, the initial technology is not good enough and leads to the appearance of some disappointed adopters who then disclose their negative reports on the public information platform, and every other enterprise refers to their negative experience. Given that SMEs are assumed to be risk-aversive in this model, it naturally causes the boom of rejecters. In addition, because the sales of EETs are rather limited and the technologies are underdeveloped, which makes it difficult to satisfy adopters. At this stage, the whole system is full of disappointed adopters and rejecters. It seems to  be impossible to reverse the bad situation, but actually this state does not last permanently for two reasons. The first reason is that this platform does not allow any rejecters to disseminate any negative information. Instead, only a handful of disappointed adopters keep other enterprises from adopting the EETs. The cascade of rejecters paradoxically prevent enterprises from being disappointed adopters who are actual bad news producers. Hence, rejecter plateau is not a stable status. Second, although during the period of rejecter plateau the inter-firm influence in the entire system is completely negative, the external influence, which is always positive, constantly exists and impacts the non-adopters (who are indeed the majority). Although the probability for a rejecter to become an adopter is fairly low, it is possible that some of the rejecters would turn into adopters under the external influence as time goes on, considering that the number of non-adopters is very large. Once there appears some satisfied  adopters, positive influence can use the “$q_s$” to diffuse, and the technologies begin to revive, which then facilitates satisfying adopters. However, in a few of previous runs, rejecter plateaus do not collapse in a meaningful period of time, as (d) of Figure \ref{fig:rejecters_plateau} shows. Once a rejecter plateau is formed, the whole diffusion process is delayed. The information platform is a double-edged sword which would either accelerate the diffusion of EETs or delay it. As analyzed above, the existence of rejecter plateau is the downside of the information platform where enterprises having bad experience of using EETs post their negative information, which may delay the whole process of technological diffusion. It is reasonable to infer that the initial number of disappointed adopters should be responsible for this phenomenon. The following experiment confirms that the initially inferior performance of a technology, which may cause a lot of disappointed adopters, significantly impacts the diffusion pattern of EETs.

\subsection{Experiment 2: the impact of initial technology performance} 
\begin{figure}[H]
	\includegraphics[width=.33\textwidth]{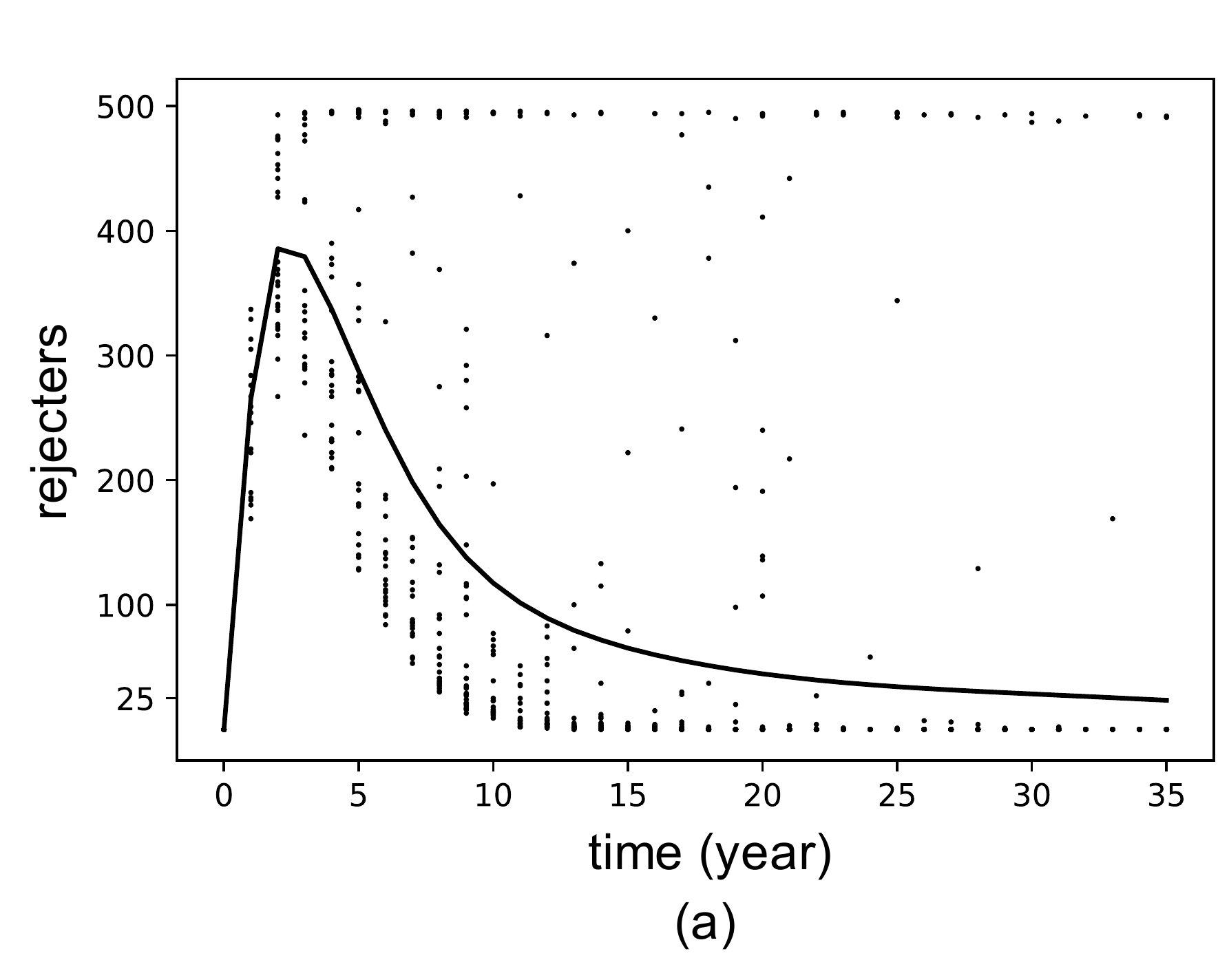}\hfill 
	\includegraphics[width=.34\textwidth]{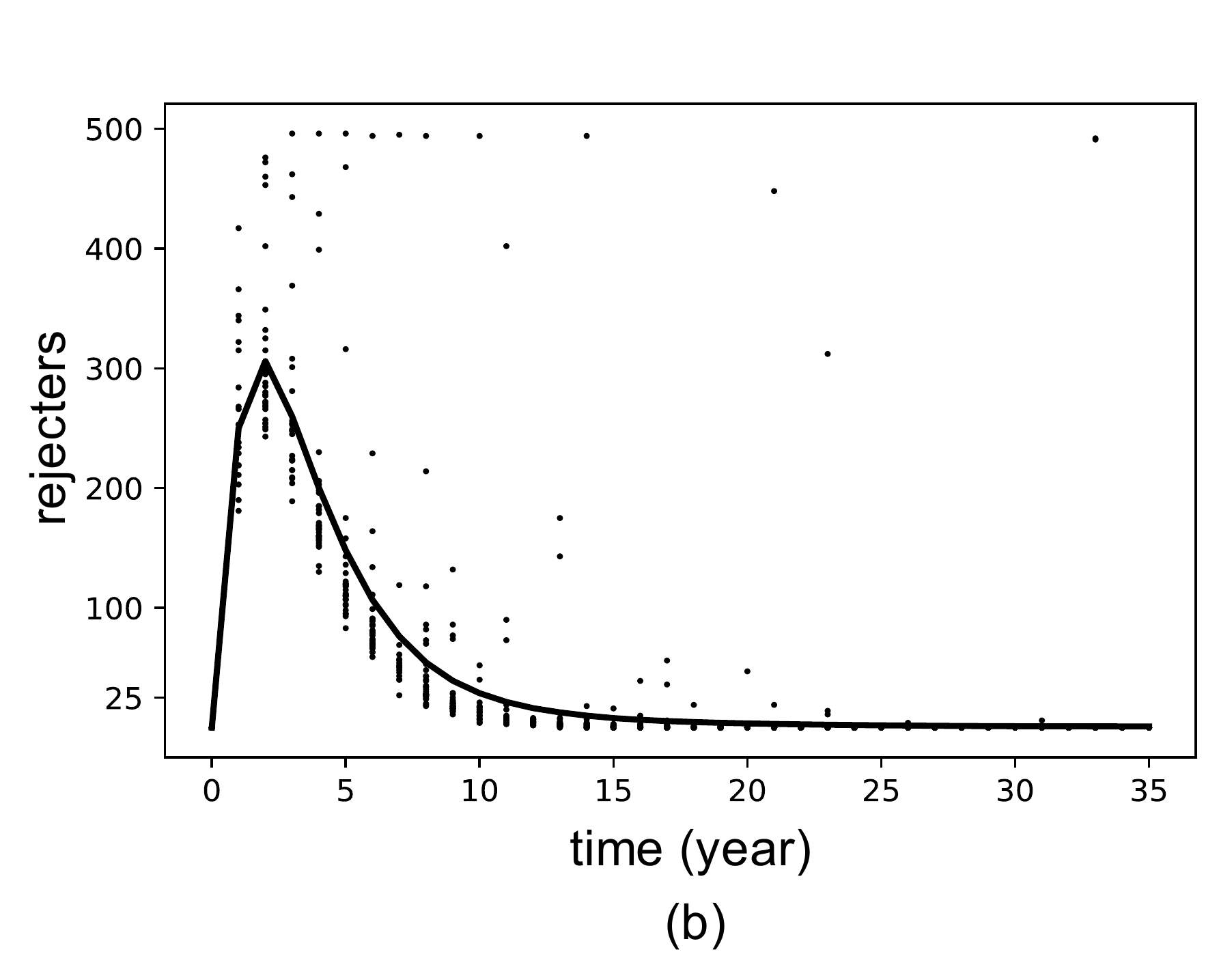}\hfill 
	\includegraphics[width=.33\textwidth]{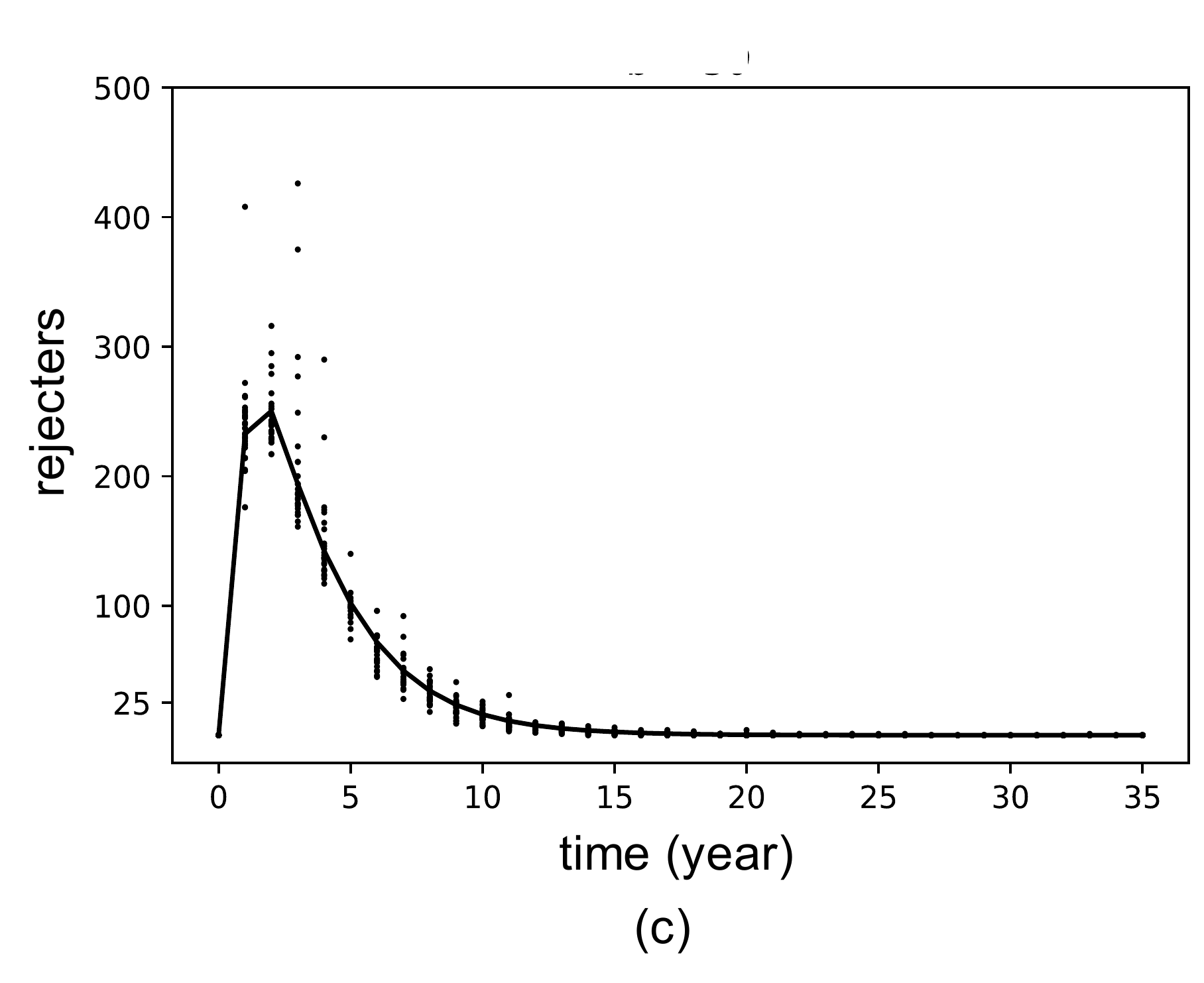}		 
	\centering
	\caption{Experiment 2 --- The number of rejecters with different values of (a) $b=10$ (b) $b=30$ (c) $b=50$}
	\label{rejecterswithbs}
	
\end{figure}

\begin{figure}
	\centering
	\includegraphics[scale= 0.6]{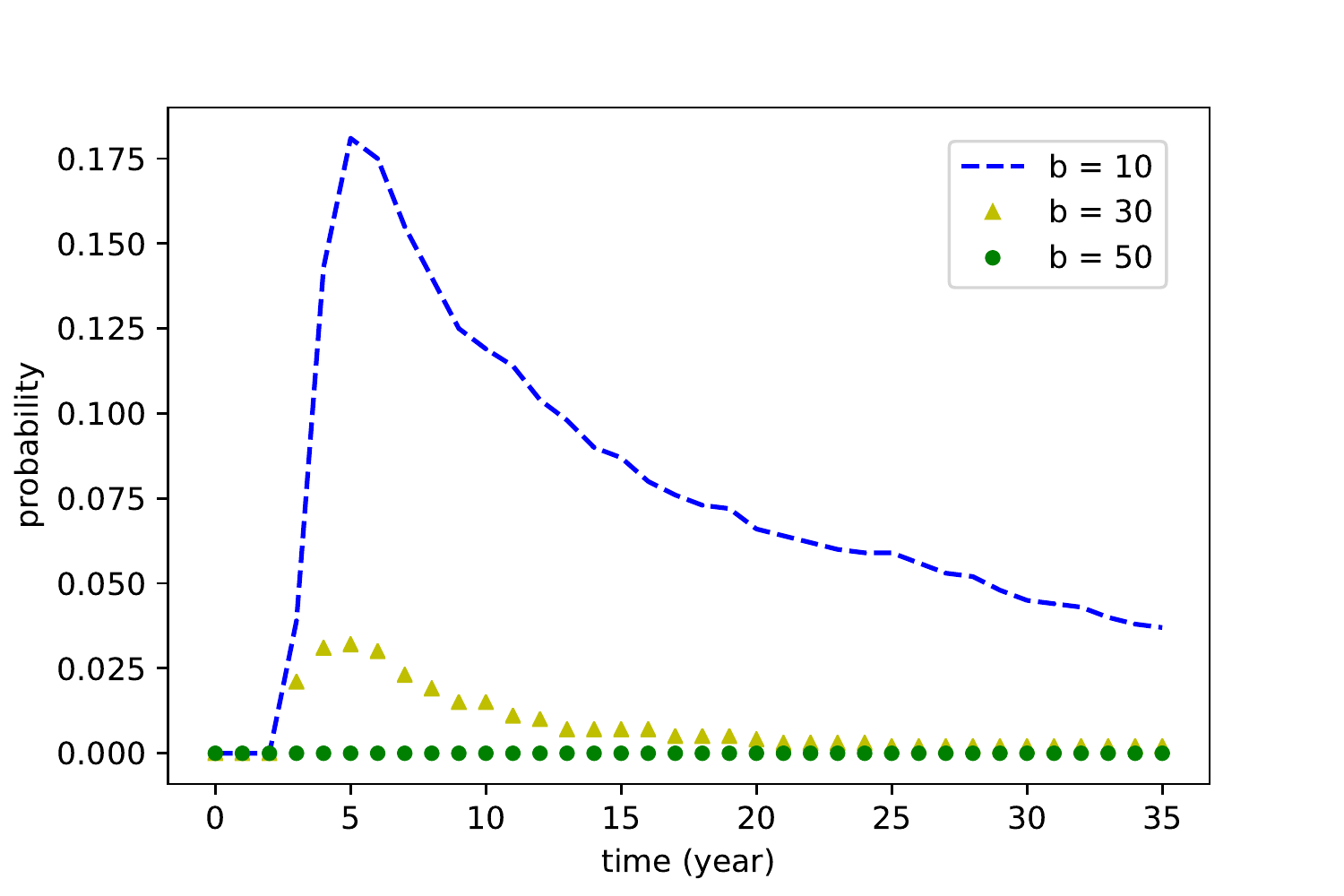}
	\caption{Experiment 2 --- The probability of rejecter plateau of each year}
	\label{Rejct_plateau_probability}
\end{figure}

When disappointed adopters, instead of satisfied ones, appear first, the system's unpredictability increases. In order to find the statistical regularity of how initial technology performance could change the diffusion pattern, we ran the model under three different initial technology performances, 1000 times for each performance. Because EETs are defined as a technology vector $\vec{T}=[T_s,T_d]$, where $T_s$ represents the static attributes of a technology, which is fixed at 40, while $T_d$ stands for a technology's dynamic attributes correlated with technological progress. When $t=0$ there is no adopter in the market, thus $T_d = b$, which represents the initial performance of a technology. The model has been run when $b=10$, $b=30$, and $b=50$, representing low, medium, and high initial technological performance respectively. Figure \ref{rejecterswithbs} shows the results of the runs with different values of $b$. The scattered dots are randomly sampled from the original data, and the solid curve is the mean values of the number of rejecters in each year. Two obvious trend can be observed: when $b$ becomes greater, the number of rejecters becomes more concentrated around the curve of mean values, and the mean value of rejecters in each year decreases. Figure \ref{Rejct_plateau_probability} shows the probability of the existence of rejecter plateaus in each year. Herein, a rejecter plateau is defined as the situation when the whole system consists of only disappointed adopters and rejecters, i.e., number of rejecters $+$ number of disappointed adopters $= 500$. As the figure demonstrates, the initial performance of a technology is responsible for the rejecter plateau. When $b$ equals 50, rejecter plateau becomes almost impossible.

A direct consequence caused by inferior technologies is the emergence of disappointed adopters. In order to examine how fragile or robust this fully-connected system is, Experiment 3 has been conducted.

\subsection{Experiment 3: the impact of disappointed adopters}
\begin{figure}[H]
	\centering
	\includegraphics[scale= 0.8]{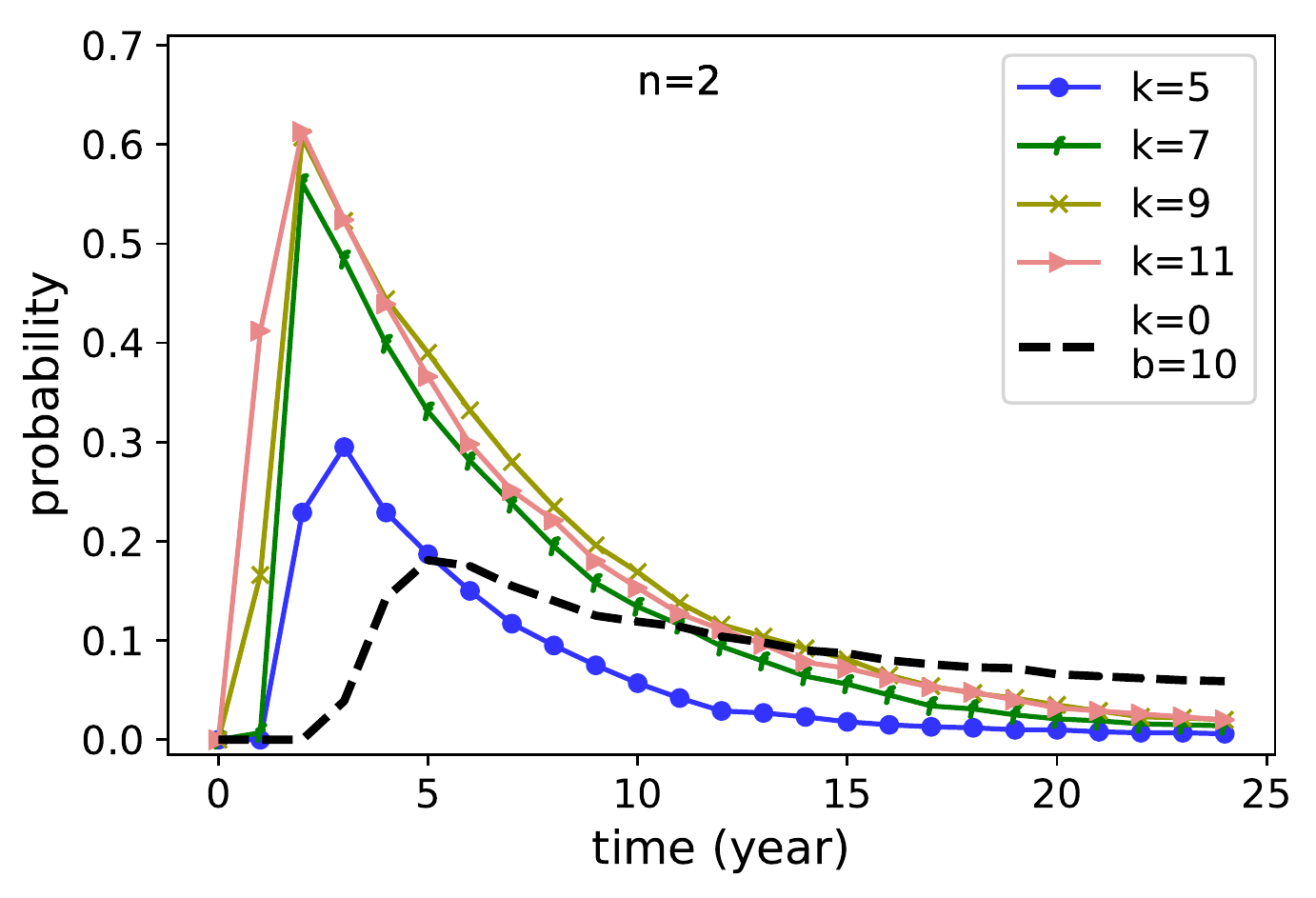}
	\caption{Experiment 3 --- The probability of rejecter plateau caused by different values of $k$}
	\label{n=2}
\end{figure}

The model has run for another 1000 times when $b$ is set as 50 (meaning the initial performance of EETs is high), which should leave little chance for the existence of rejecter plateau as Figure \ref{Rejct_plateau_probability} shows. However, a few disappointed adopters are manually added at the beginning ($0th$ year) of each run to observe how the system is affected by the initial negative information, and Figure \ref{n=2} 
is obtained. The result shows the system manifests great sensitivity to negative information. Even the initial technological performance is good enough ($b=50$), with the existence of disappointed adopters (the number of whom is denoted as $k$) the probability of rejecter plateau spikes rapidly. When $k=5$ the highest probability exceeds the probability of that when $k=0$ and $b=10$, as the black dashed line shows in Figure \ref{n=2}. An interesting subtlety is that the black dashed line peaks later than its triangle-like companions', which means rejecter plateaus induced by inferior technologies need more time to foster its disappointed adopters. Another prominent contrast is that the black dashed line has a much heavier tail than the other graphs in Figure \ref{n=2}, which suggests the rejecter plateaus caused by “fake” disappointed adopters are much more fragile than those caused by real inferior technologies.
%\newpage

\begin{figure}[H]
	\centering
	\includegraphics[scale= 0.6]{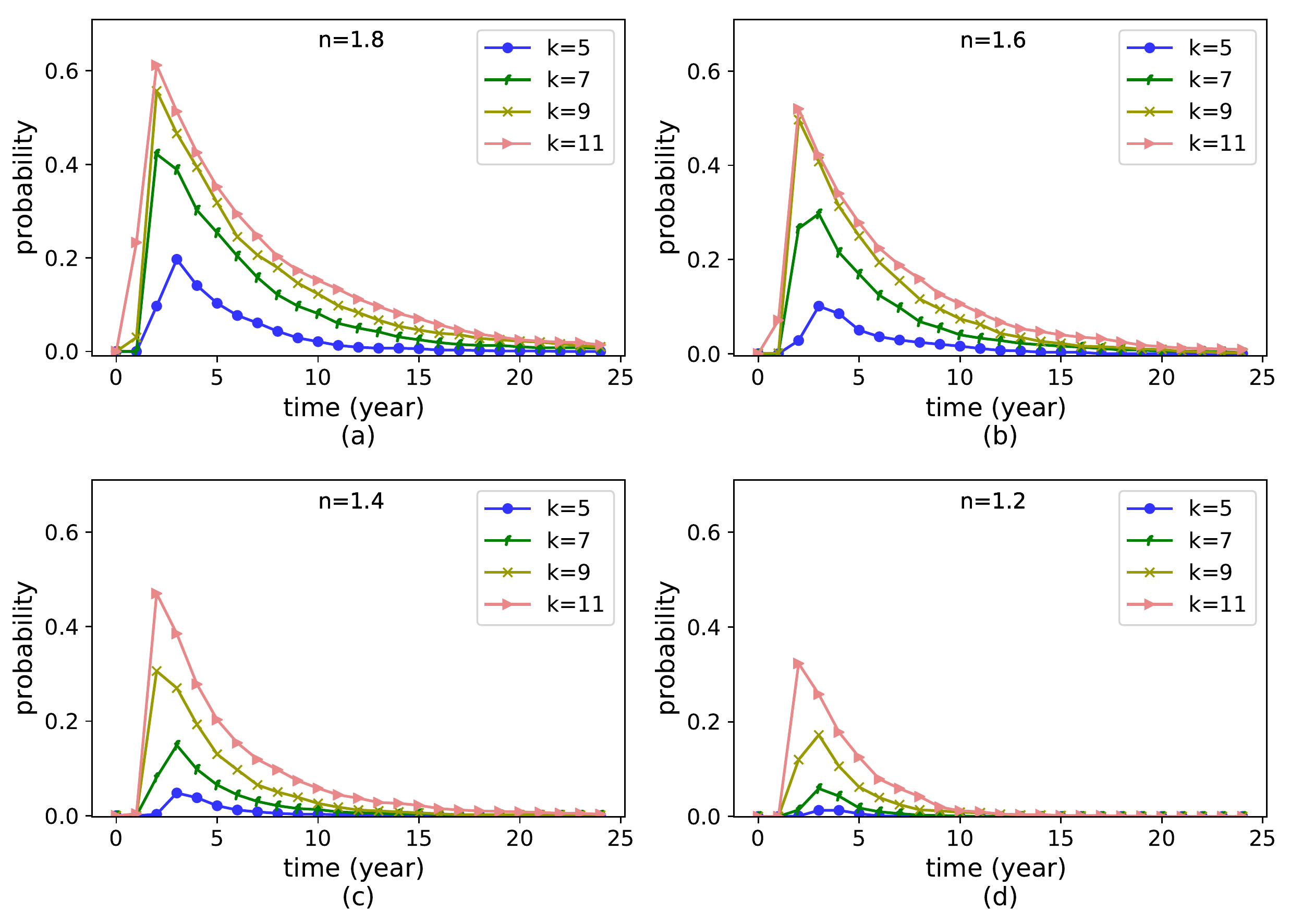}
	\caption{Experiment 3 --- The probability of rejecter plateau with different values of $k$ and $n$}
	\label{multiple_n}
\end{figure}

As Eq. (\ref{eq:negative}) indicates, another critical factor responsible for the model's extreme behavior is $n$, which adjusts the degree of enterprises' risk aversiveness, is a critical factor that differs negative information from positive information. The $n$ boosts the transmission of negative information which turns non-adopters to rejecters and further hampers the rejecters from changing to adopters. Thus, it can be inferred that if $n$ becomes smaller, the existence of rejecter plateau in each year diminishes. Based on this inference, the model has been run for another several thousand times with different values of $n$, as shown in Figure \ref{multiple_n}. Notably, reducing $n$ can decrease the probability of rejecter plateau's existence. 

\subsection{Experiment 4: network density ($nd$) and interactive coefficient ($q_s$)}
\begin{figure}[H]
	\centering
	\includegraphics[width=0.8\textwidth]{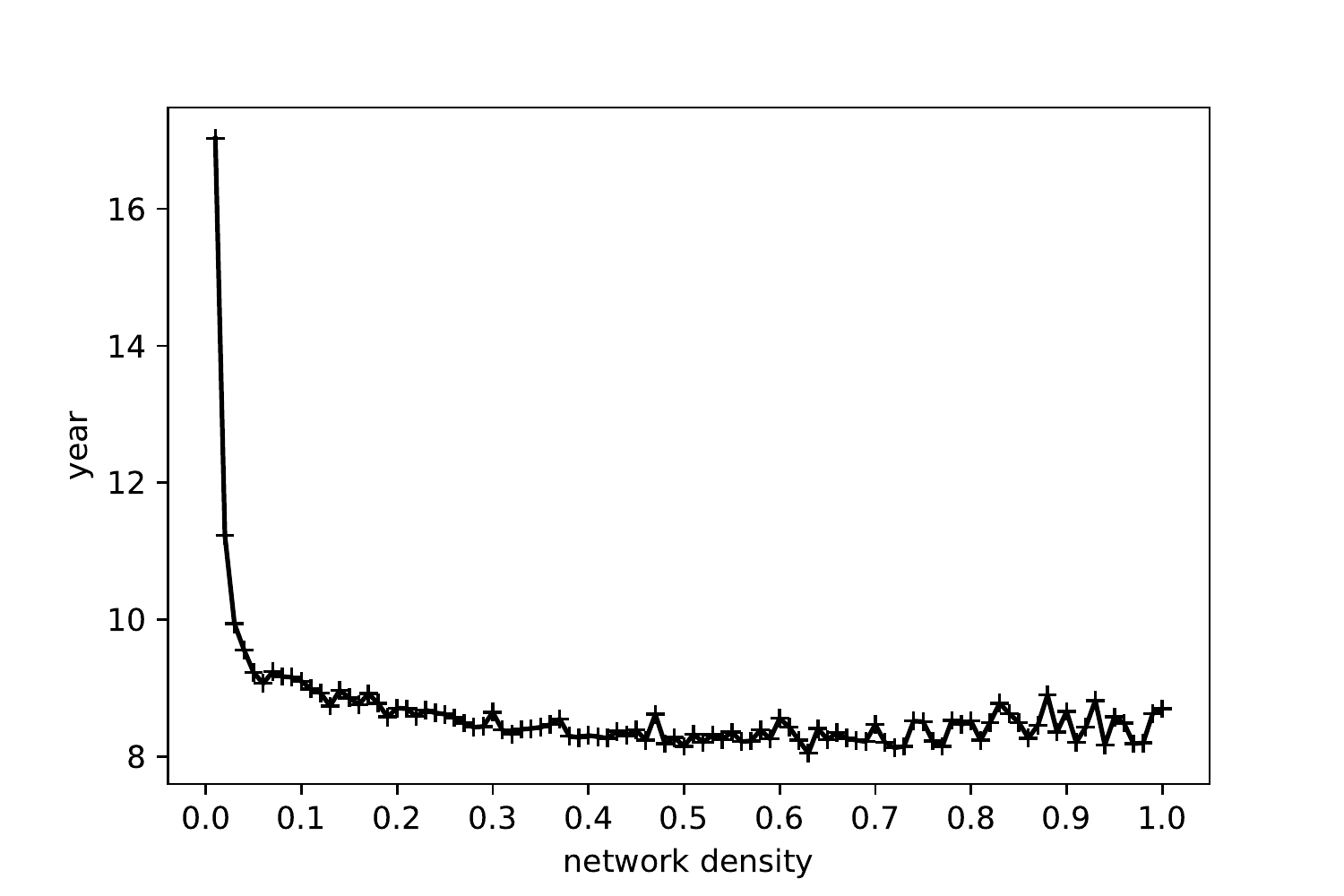}
	\caption{Experiment 4 --- The impact of network density on EET diffusion}
	\label{fig:linkTest}
\end{figure}

The above analysis shows that when network density becomes 1, the system is highly sensitive to negative information and may cause a phenomenon called rejecter plateau which postpones massive adoption. It is natural to suppose that the high network density accelerates the diffusion but also makes the whole system extremely volatile. However, when we take a closer observation on how network density impacts innovation diffusion, a counter-intuitive pattern emerges. The model has been run while the network density varies from 0.01 to 1 with even intervals of 0.01, 300 times for each value, with other parameters remain the same as those of Experiment 1. Figure \ref{fig:linkTest} is obtained, with the vertical axis representing the mean values of the $year_{0.95}$ which denotes the year when EET diffusion reaches 95\% of market share, and the horizontal axis representing network density ($nd$). As can been seen, network density has significant impact on EET diffusion when the system's network density is very low. The $year_{0.95}$ declines drastically from 17 to 9 when network density changes from 0.01 to 0.05; but then it stays between 8 and 9 no matter how the network density changes. 

The first glance at this pattern might cause confusion, but the explanation resides in the assumptions made above. It is noteworthy that network density essentially determines the number of neighbors of each agent. Eq. (\ref{eq:positive}) and (\ref{eq:negative}) are assumed to determine the degree of how intensive an agent is informed about good or bad news. These two equations basically are exponential function tending towards 1 when the powers tend to be infinitely great. Specifically, according to the two equations, there is no great difference between 25 and 100 or more influential neighbors. This is consistent with the diminishing marginal effect in economics and the empirical observation in psychology that when the number of influential neighbors who share the same opinion exceeds a certain number (e.g. 4), a human agent is very much likely to be convinced \cite{asch1955,myers2012,sznajd2014}, which should also make sense for enterprises, whose components are people. This also can be explained with common sense: one only needs \textit{a few of} or \textit{enough} friends to refer to before buying something, instead all of them. Last but not least, according to the random graph theory, a random network has a critical network density, beyond which the probability of the emergence of a giant component (a big connected fraction of the original graph) is extremely high \cite{}. It is intuitive that innovations diffuse much faster in a connected network than they do so in graphs including isolated nodes. These microscopic assumptions and facts cause the emergence of Figure \ref{fig:linkTest}, which is defined as “saturation effect” because the impact of increasing network density on the speed of EET diffusion tends to be saturate.

\begin{figure}[H]
	\centering
	\includegraphics[scale=0.55]{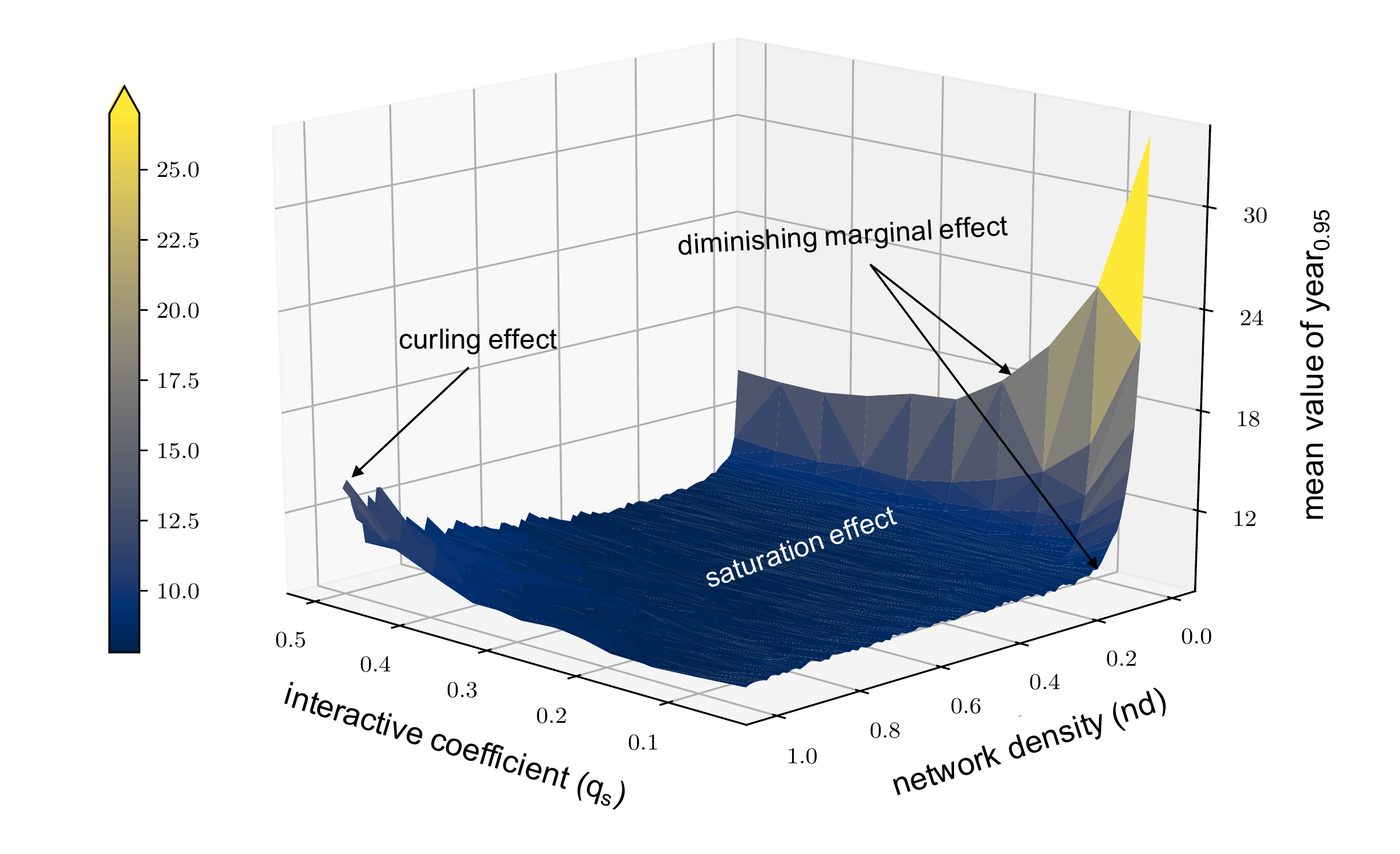}
	\caption{Experiment 4 --- The combined influence of network density and interactive coefficient on system efficiency}
	\label{fig:nd_q_mean}
\end{figure}

\begin{figure}[H]
	\centering
	\includegraphics[scale=0.5]{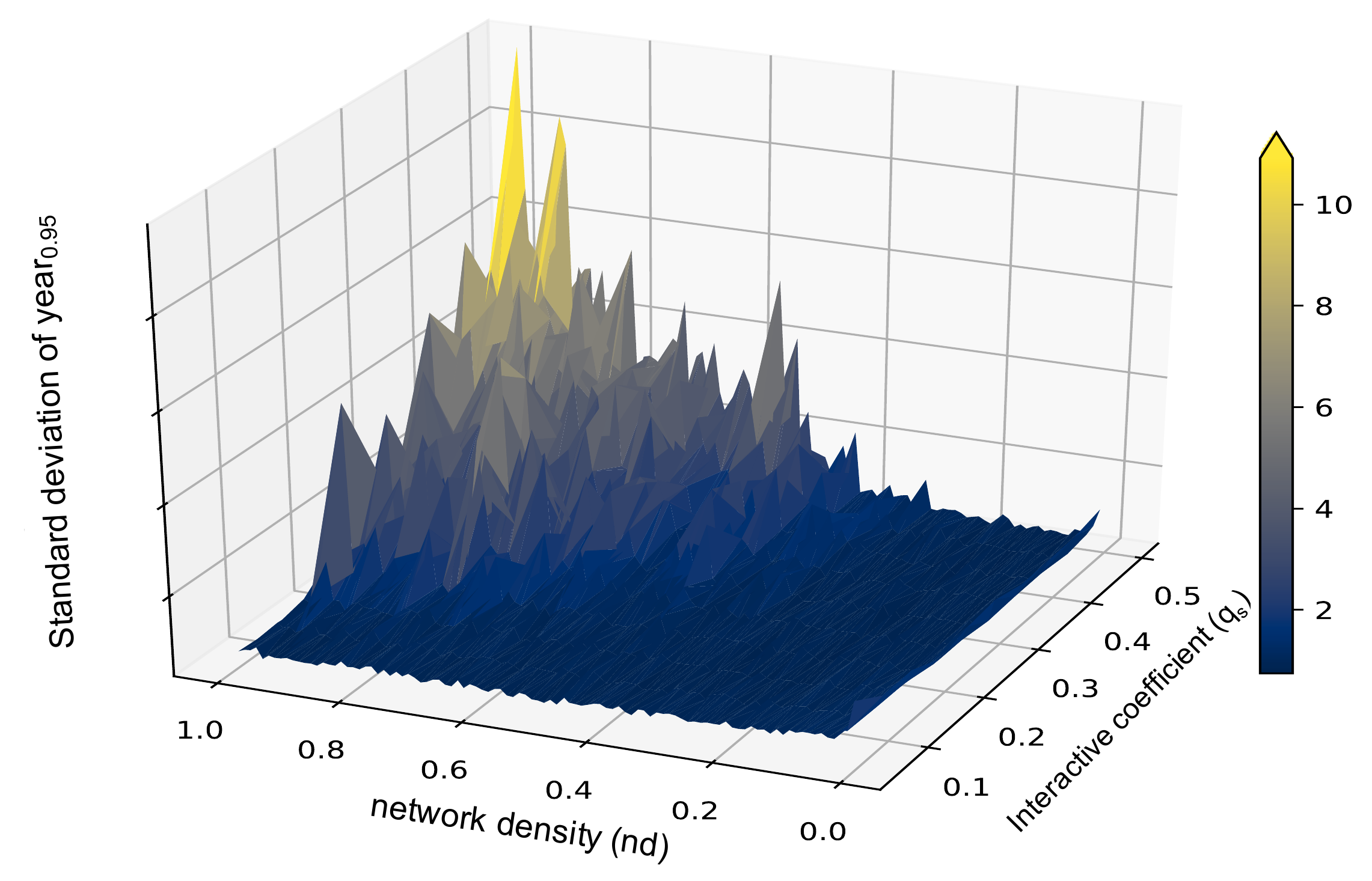}
	\caption{Experiment 4 --- The combined influence of network density and interactive coefficient on system stability}
	\label{fig:nd_q}
\end{figure}

It should not be ignored that the interactive coefficient ($q_s$) in Eq. (\ref{eq:positive}) and (\ref{eq:negative}) is another critical factor influencing the diffusion speed. According to similar reasons, the impact of increasing $q_s$ also diminishes. To visualize the combined the impact of network density and interactive coefficient, we run the model for 100,000 times in total with $nd$ changing from 0.01 to 1 and with $q_s$ changing from 0.05 to 0.5 with even intervals of 0.01 and 0.05 respectively. Figure \ref{fig:nd_q_mean} is derived. It is clear that when interactive coefficient and network density are very small ($q_s$ $<$ 0.15 and $nd$ $<$ 0.05), increasing them can significantly accelerate EET diffusion. But their effects are not limitless, diminishing marginal effect of both interactive coefficient and network density gives rise to the saturation effect. As shown in Figure \ref{fig:nd_q_mean}, the wide and flat bottom illustrates the saturation effect. Within the saturation domain, the impact of the variation of $q_s$ and $nd$ is negligible. However, this flat area results in curling up when $q_s$ and $nd$ tend to 0.5 and 1, which is labeled as "curling effect". It is reasonable to recall the rejecter plateau phenomenon and attribute the curling to the system's excessive volatility caused by strong and frequent inter-firm influence at the microscopic level. This is confirmed by Figure \ref{fig:nd_q}, whose bottom axes stay the same while the vertical axis is changed to represent the standard deviation(SD) of $year_{0.95}$. Obviously, the diffusion becomes more unpredictable when interactive coefficient ($q_s$) and network density ($nd$) increase. Particularly, a few of the runs generate the SD of $year_{0.95}$ greater than 12. These runs may result in $9+12=21$ years, which are obviously larger than a “normal” 17 years in Experiment 1.
\section{Discussion}
Inspired by the Bass model, the influence between adopters and potential adopters is very powerful to shape the diffusion pattern of innovations, which could be exploited in the diffusion of EETs among SMEs. Although the inter-firm influence is constrained due to firms' lack of motivation to share information and the lack of observability of EETs, an information platform together with proper regulation could activate the information-disclosure behavior of enterprises, given that many EETs are not close to their core business and disclosing relevant information would not harm their core competitiveness. In this context, this study aims to explore whether and how such an information-platform would change the diffusion pattern of EETs among SMEs. 

Turning back to the issues mentioned in the Introduction, the existence of an information platform where SMEs can post their experience of using EETs changes the diffusion pattern of EETs both quantitatively and qualitatively, as the result illustrated in Experiment 1. The diffusion does not have a thin tail in the first few years, but takes off immediately, which indicates the effectiveness of such an information platform. However, not every run can generate this pattern. Because of the existence of negative information, the completely connected system becomes an “accomplice” to facilitate the diffusion of “bad news”, especially when the potential adopters are risk-aversive. As analyzed above, this system is sensitive to initial conditions, and rejecter plateaus are caused by the disappointed adopters at the beginning of each run, which leads to the consideration of the initial performance of EETs. 

Experiment 2 demonstrates that the initial performance of EETs is a critical factor responsible for rejecter plateaus, which implies that the companies who try to sell their EETs should be very careful for launching their immature products in a system whose components are fully connected. Otherwise, the early adopters would be disappointed and disseminate negative information that could rapidly foster a large number of rejecters and delay massive adoptions. Although, the technological progress simulated in our model is not slow: $a=2.5$ means one adopter contributes to 2.5 units of technological progress; for an innovation reaching 95\% market share in the 17th year, it becomes fully mature approximately in the 10th year. Once an EET is evaluated as inferior by the disappointed adopters at the beginning, the companies who sell this EET
 may lose the chance to gain revenue to improve the technologies. Therefore, deliberate measures should be formulated when promoting promising EETs with inferior initial performances. Wang et al. \cite{WANG2018} suggested that information screening could be a useful way to protect the diffusion of PV systems from rumors (e.g., harmful radiation), which could also be used to maintain the information platform. Other measures such as to appropriately postpone the display of negative information submitted by enterprises may also be an effective way to let promising EETs show their goodness. It is noteworthy that extra experiments manifest that even with bad initial performance, promising technologies can evolve more smoothly when network density is much lower than 1, which suggests that infant EETs need time to grow up and thus need protection.

Experiment 3 investigates how disappointed adopters at the beginning impact the occurrence frequency of rejecter plateaus. The results show that the more disappointed adopters appear at the beginning, the more likely rejecter plateau would occur. Moreover, we have also tested how $n$ representing the risk attitude of SMEs affects the EET diffusion. The result shows a prominent decreasing trend with the decline of $n$, which indicates the degree of risk-aversion significantly influences the probability of rejecter plateau's occurrence. When $n\rightarrow1$ and $k\rightarrow0$, the occurrence probability of rejecter plateau tends to 0, which allows the technology to manifest its normal performance (note the initial technological performance is set at high level), instead of being undervalued.  Policymakers could leverage the parameter $n$ by effective policies to reduce the risks perceived by enterprises, especially for promoting promising technologies with humble beginnings. Financial incentives and risk compensation are typical economic policies that are aimed at improving SMEs' willingness to pay for EETs through reducing enterprises' perceived risks. They are particularly suitable for EETs with considerable positive externalities but high risks which have great promotion necessities \cite{CAO2016}. Moreover, demonstration projects of EETs could be launched before commercialization because it leaves a good first impression of EETs for potential adopters. Positive information from demonstration projects could prevent the dominance of initial disappointed adopters, thus suppress the occurrence of rejecter plateaus. 

\begin{figure}[H]
	\centering
	\includegraphics[scale=0.5]{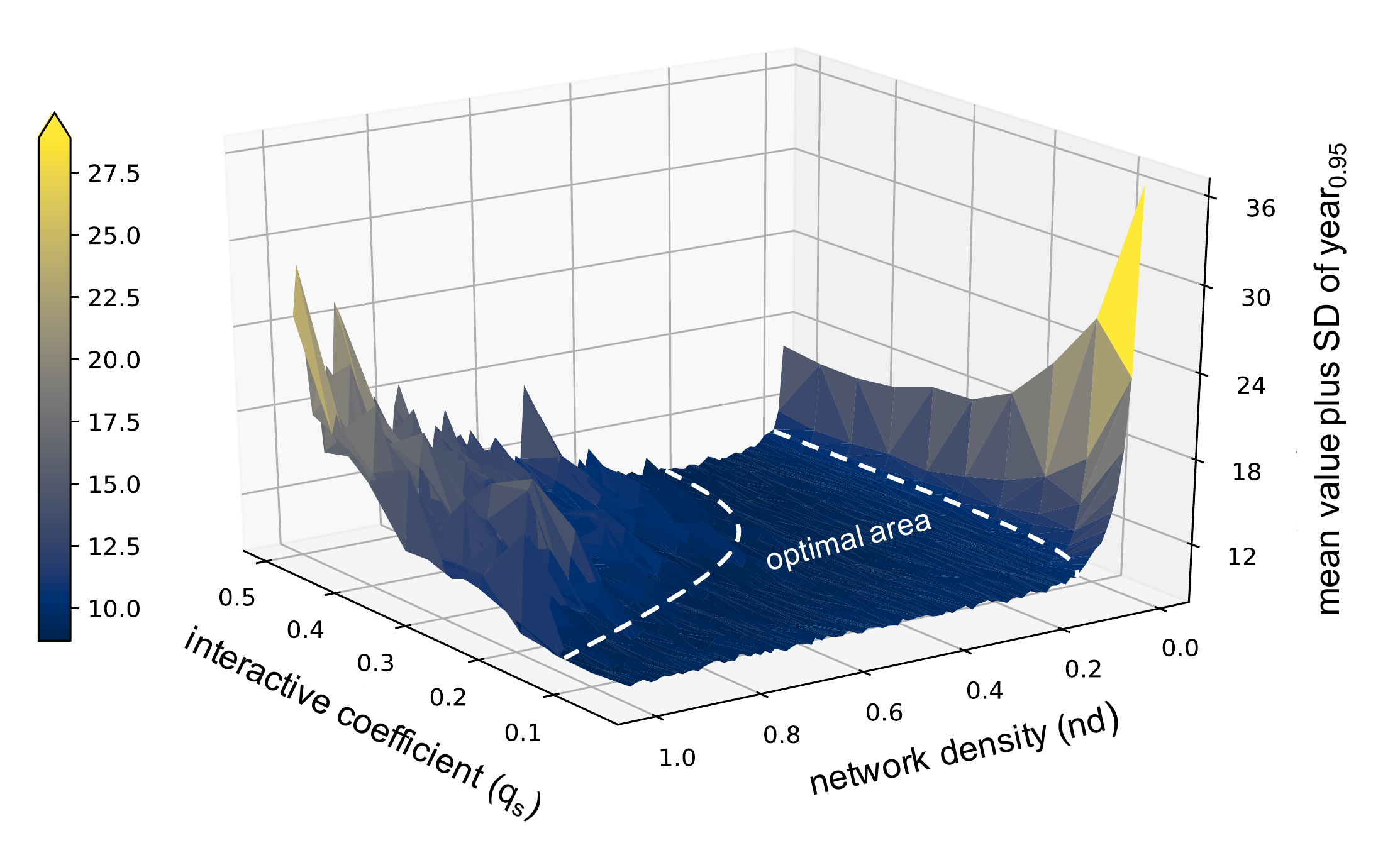}
	\caption{The mean value plus SD of $ year_{0.95}$ influenced by network density and interactive coefficient}
	\label{fig:nd_q_mean_SD}
\end{figure}

Experiment 4 examines how network density and interactive coefficient impact the diffusion rate of EETs. The results display a significant reduction of $year_{0.95}$ when interactive coefficient and network density increase. However, their performances diminish dramatically after reaching a certain value (e.g. $q_s$ = 0.15, $nd$ = 0.05), then impose negligible influence on the diffusion rate, which is labeled as “saturation effect”. The saturation effect is eventually terminated by the “curling effect”, which indicates that a system with intensive internal influence between its components could be excessively volatile and harmful to EET diffusion. Thus, making trade-off between stability and diffusion speed is crucial. By combining Figure \ref{fig:nd_q_mean} and \ref{fig:nd_q}, we can obtain Figure \ref{fig:nd_q_mean_SD}, which has a vertical axis representing the mean value of $ year_{0.95}$ plus its standard deviation. Apparently, the flat deep blue area marked as “optimal area” consists of the combinations of different values of network density and interactive coefficient, which maximize the diffusion rate of EETs as well as minimize the instability. However, the current network density estimated in this paper is near 0.01. A lower estimation would also be more appropriate given that EETs are not closely related to the majority of enterprises' core processes. Additionally, the lack of observability of EETs makes information flows more difficult. When the network density is rather low, it is quite effective to speed EET diffusion rate by adding more connections between the adopters and potential adopters. Building formal or informal EET-related communities, associations, and organizations incorporating enterprises could be effective measures to enhance their EET-related communications, then increase EET diffusion rate.
Nevertheless, it is unrealistic to find a measure to precisely manipulate the network density so that the system can locate at any spot in the optimal area. Building an EET information platform can increase network density but also leads to the other extreme ---  excessive volatility, which is harmful to the diffusion of EETs. Fortunately, Figure \ref{fig:nd_q_mean_SD} shows there is a small portion of optimal area accessible even when the network density is 1. Numerically, when $q_s$ locates between 0.1 and 0.15 the curling effect is significantly mitigated. This implies that EETs would diffuse more smoothly, if the enterprises are less inter-dependent. Therefore, a great emphasis should be placed on the orientation of the EET information platform: it should be informative rather than judgmental. The information platform should be a source of useful information that can help enterprises search and select suitable EETs for themselves, instead of a place for advertising or pouring complaints. A structured form properly confining what and how information should be disclosed could be helpful and might be developed in future research.
\section{Conclusions and policy implications}

Improving energy efficiency is widely regarded as the key to combat the energy dilemma and climate change \cite{shao2019can}. Promoting the diffusion of energy efficiency technologies (EETs) among small and medium-sized enterprises is a very promising approach to achieve this goal. However, the diffusion rate of EETs is still far from satisfactory. Inspired by the innovation diffusion theory \cite{Rogers1962} and the Bass diffusion model \cite{BASSFRANK1969}, this paper proposes that leveraging the inter-firm influence can effectively stimulate EET diffusion. Considering the EETs are normally lack of observability and distant from enterprises' core businesses, building an information platform where enterprises share their experience of adopted EETs is a feasible measure to activate the inter-firm influence. This paper builds an agent-based model and conducts a series of numerical experiments to explore the impact of such information platform on EET diffusion. Concretely, this paper investigated three questions: (1) Can this information platform accelerate EET diffusion? (2) Will negative reports posted on the platform hamper EET diffusion? (3) If the answer is positive, how should negative information be handled by system designers or policymakers? Answering these questions can help policymakers proactively make policies conducive to maintaining a robust information platform that promotes EET diffusion. EET suppliers can also obtain useful information from the findings to strategically avoid failures caused by the intensive interactions between enterprises.

The information platform significantly densifies the network that connects the enterprises. The first numerical experiment intuitively demonstrates that the information platform considerably accelerates EET diffusion. However, the dense network also boosts negative information to diffuse even faster, as the enterprises are assumed risk-aversive. The rapid encroachment of negative information eventually delays the massive diffusion of EETs. Particularly, when an EET is perceived as inferior by adopters in the early stage of diffusion, a large number of rejecters would emerge and aggregate as a "rejecter plateau" resisting EET diffusion. This finding implies that EET suppliers should be very cautious about launching their immature products. Authorities could assist EET suppliers in seeding positive impressions on the market through demonstration projects. The second numerical experiment shows that “fake disappointed adopters” can also trigger the surge of rejecters. Therefore, information screening might be a necessary intervention by authorities to eliminate rumors and maintain a robust information platform. The third numerical experiment shows that enterprises' sensitivity to risks is a critical factor responsible for the diffusion system's reactions to negative information: lowering the perceived risks effectively alleviates the surge of rejecters under the influence of negative information. Subsidization might be effective to compensate for the expected loss of potential EET adopters and thereby reduce their perception of risks. The fourth numerical experiment indicates that a moderately low intensity of inter-firm interaction could mitigate the excessive volatility caused by network densification. Thus, policymakers should constrain the way of enterprises' information disclosure behavior lest the information platform becomes judgmental rather than informative. A structured form that clearly defines what information should be provided might lead enterprises to rationally share and comprehend the disclosed information.

In conclusion, an information platform can activate the inter-firm influence and stimulate the diffusion of EETs among enterprises; however, the downside of the information platform should be noticed and carefully controlled.

\section*{Acknowledgements}
This work is supported by the National Natural Science Foundation of China (No. 71673023) and China Scholarship Council ([2017]3109). In addition, we would like to convey special thanks to Prakash Reddy Thimmapuram for his very useful and constructive suggestions on our work.

\section*{Conflicts of interest}
The authors declare no conflict of interest.

\bibliographystyle{elsarticle-num}
\bibliography{References}

\end{document}